\documentclass{IEEEtaes}
\usepackage{cite}
\usepackage{amssymb,amsmath,latexsym,amsfonts,amsthm,mathtools}
\usepackage{multirow}

\usepackage{graphicx}
\usepackage{cite}
\usepackage{float,subcaption}
\usepackage{textcomp}
\usepackage{xcolor}
\definecolor{darkpastelpurple}{rgb}{0.59, 0.44, 0.84}
\usepackage{hyperref}
\hypersetup{colorlinks, breaklinks, citecolor=blue, linkcolor=blue, urlcolor=blue}
\usepackage[nameinlink]{cleveref}
\Crefformat{figure}{#2Fig.~#1#3}
\Crefmultiformat{figure}{Figs.~#2#1#3}{ and~#2#1#3}{, #2#1#3}{ and~#2#1#3}
\crefmultiformat{figure}{Figures~#2#1#3}{ and~#2#1#3}{, #2#1#3}{ and~#2#1#3}

\usepackage{tikz}
\usetikzlibrary{automata, shapes, arrows, calc, arrows.meta, fit, positioning}

\theoremstyle{plain}
\newtheorem{theorem}{Theorem}
\newtheorem{lemma}{Lemma}
\newtheorem{corollary}{Corollary}

\newtheorem*{problem*}{Problem}
\theoremstyle{remark}
\newtheorem{remark}{Remark}
\newtheorem{assumption}{Assumption}

\newtheorem{definition}{Definition}
\theoremstyle{definition}

\newcommand{\sign}{\mathrm{sign}}

\newcommand{\sgmf}{\mathrm{sgmf}}
\newcommand{\m}{\mathrm{m}}
\newcommand{\M}{\mathrm{M}}

\usepackage{pgfplots}

\usepackage{mathrsfs}
\usepackage{cuted}
\usepackage{lipsum,multicol}

\jvol{XX}
\jnum{XX}
\jmonth{XXXXX}
\paper{1234567}
\pubyear{2025}
\doiinfo{TAES.2025.Doi Number}
\allowdisplaybreaks
\IEEEoverridecommandlockouts
\begin{document}

	\title{Time-Constrained Interception of Seeker Equipped
		Interceptors with Bounded Input}

	\author{Ashok Samrat R}
	
	\affil{Indian Institute of Technology Bombay, Powai, Mumbai 400076, India} 
	
	\author{Swati Singh}
	\member{Graduate Student Member, IEEE}
	\affil{Indian Institute of Technology Bombay, Powai, Mumbai 400076, India} 
	
	\author{Shashi Ranjan Kumar}
	\member{Senior Member, IEEE}
	\affil{Indian Institute of Technology Bombay, Powai, Mumbai 400076, India}

	
\accepteddate{XXXXX XX XXXX}
\publisheddate{XXXXX XX XXXX}

\corresp{{\itshape (Corresponding author: S. Singh)}}

\authoraddress{The authors are with the Intelligent Systems and Control Lab, Department of Aerospace Engineering, Indian Institute of Technology Bombay, Powai, Mumbai 400076, India. (e-mails: \href{22m0019@iitb.ac.in}{22m0019@iitb.ac.in},\href{swatisingh@aero.iitb.ac.in}{swatisingh@aero.iitb.ac.in}, \href{srk@aero.iitb.ac.in}{srk@aero.iitb.ac.in}).}


\maketitle
\begin{abstract}
	This paper presents a nonlinear guidance scheme designed to achieve precise interception of stationary targets at a pre-specified impact time. The proposed strategy essentially accounts for the constraints imposed by the interceptor's seeker field-of-view (FOV) and actuator limitations, which, if ignored, can degrade guidance performance. To address these challenges, the guidance law incorporates known actuator bounds directly into its design, thereby improving overall interceptor effectiveness. The proposed method utilizes an input-affine magnitude saturation model to effectively enforce these constraints. By appending this input saturation model to the interceptor's kinematic equations, a guidance law is derived that ensures interception at the desired impact time while accounting for the physical constraints of the sensor and actuator. The efficacy of the proposed strategies is demonstrated through comprehensive numerical simulations across various scenarios and is compared against an existing guidance strategy.
\end{abstract}
\section{Introduction}
\IEEEPARstart{A} successful salvo attack \cite {11018241} relies on the precise interception of the target at the desired time \cite{jeon2006impact}. Most terminal-constrained guidance laws achieve interception with zero miss distance \cite{8070985}. However, their effectiveness is often limited by the physical constraints of the interceptor, particularly field-of-view (FOV) and actuator saturation. The seeker’s FOV bound determines the angular range within which an interceptor's seeker or sensor can detect and track a target; however, saturation of lateral accelerations imposes a critical in-flight constraint that affects the ability to meet terminal constraints requirements, as it restricts the rate of lead angle variation. Addressing these constraints is essential for enhancing guidance law performance and ensuring target acquisition despite actuator limitations.

While the impact time-constrained guidance (ITCG) has been extensively studied, few works have incorporated seeker FOV constraints. Guidance strategies were also proposed using sliding mode control to satisfy impact time constraints proposed in \cite{SRK2015largeHE,9000526}. Output shaping and heading angle shaping impact time guidance was designed in \cite{doi:10.2514/1.G008834,10994241}. Another heading error, shaping-based guidance, was designed in \cite{10419004}. The authors in \cite{doi:10.2514/1.G004074} proposed a generalized circular impact-time guidance strategy. In \cite{9574446}, the authors proposed a composite guidance to control the interception time while obeying the physical constraints of the interceptor. Authors in \cite{8382207} proposed a guidance scheme focusing on imposing FOV constraints subject to impact angle requirements. However, guidance laws to ensure some desired impact time while meeting FOV constraints are not as numerous.
The FOV bound was also addressed in \cite{Jeon2017FOV} and \cite{Shim2018FOV}, but the maximum achievable impact time remained constrained due to the monotonically decreasing lead angle variation. Alternatively, authors in \cite{H_J_Kim} proposed a strategy based on the backstepping technique, where a two-phase lead angle variation was used with an approximated signum function. During the first phase, the lead angle was non-zero but constant, and then it became zero in the second phase. This allowed for the easy imposition of the FOV bounds using the lead angle constraints. 

In three-dimensional (3D) engagement scenarios, the kinematic equations are highly coupled, making it challenging to directly extend guidance laws developed for planar engagements. Despite its significance, ITCG with FOV constraints in 3D scenarios has received limited attention in existing research. In \cite{Kumar2022BarLyap}, the authors used time-go-estimate to design a barrier Lyapunov-based guidance strategy that was able to achieve interception at the desired impact time. It was also shown that this guidance strategy remained nonsingular for the given range of achievable impact times. The guidance strategies were applicable to both planar and 3D engagements while considering FOV constraints. A polynomial shaping-based $3$D impact time guidance that respects the seeker's field-of-view was discussed in \cite{SURVE2022406}. In this work, the author showed that the proposed guidance strategy ensured a near-zero miss distance interception at a desired impact time for constant speed targets. In \cite{10354070}, a time-to-go estimation free guidance strategy for impact time control was presented. However, the guidance law used range-to-go estimates to achieve target interception at a desired impact time. Authors in \cite{10271537} proposed a nonlinear impact time and angle guidance strategy that enables an interceptor to engage a stationary target while satisfying field-of-view (FOV) constraints.

While interceptors possess a higher load factor compared to manned aircraft, their maximum load factor is constrained by structural integrity and aerodynamic drag. These constraints directly impact the interceptor's lateral acceleration. 
To ensure the effectiveness of the guidance law and maintain its error convergence guarantees, it is essential to incorporate acceleration saturation effects into its design. Numerous studies have addressed the challenge of input saturation while also considering terminal constraints on the field-of-view. A time-varying sliding mode control method based on a time-based generator function proposed in \cite{Wang2022IGCInpSat}, incorporating constraints on both the impact angle and the control input. This approach was then combined with the backstepping control method to design an integrated guidance and control (IGC) scheme. Authors in \cite{Swati2023BarLyap} proposed a barrier Lyapunov function (BLF) based guidance strategy that enables the interceptor to intercept a target at the desired impact time while obeying the FOV and input constraints. Alternatively, in \cite{Liu2019IGCInpSat}, the BLF was used to impose the state constraints while the input saturation was imposed via the actuator model. Although the approaches in \cite{Swati2023BarLyap, Liu2019IGCInpSat} address the impact time control problem with FOV and input constraints, they are not applicable to three-dimensional engagement scenarios.

As far as the author knows, no prior work has addressed impact time-constrained target interception for 3D engagements while simultaneously considering both the seeker’s FOV and input bounds. To fill this research gap, we propose a nonlinear guidance scheme by incorporating the input saturation model from \cite{Zou2017InpSat} into the kinematic equations. For design purposes, the lateral acceleration is treated as an additional state. The FOV constraint is accounted for in both three-dimensional and planar engagements, using backstepping approaches based on the lead angle and heading angle, respectively. In light of existing results and the above discussion, the contributions of this work are outlined as follows:
\begin{itemize}
	\item A nonlinear guidance approach that allows the interceptor to fulfill the impact time requirements while respecting the physical bounds of the interceptor's seeker's field-of-view and the actuators is proposed for both three-dimensional and planar engagement scenarios. 
	\item In contrast to \cite{Kumar2022BarLyap, Swati2023BarLyap}, where time-to-go approximations are used, the proposed guidance strategy utilizes range-to-go, making it independent of such estimations. The derivations of guidance commands do not involve linearizing or decoupling the kinematic equations. Therefore, these guidance laws remain valid for sufficiently large initial heading angle errors. To incorporate the constraints on interceptor lateral acceleration while deriving guidance command, an input saturation model has been proposed. 
	\item Unlike most approaches in the literature, where the trajectory of the interceptor depends on the initial headings, the strategy designed in this work leverages virtual control inputs to provide additional flexibility in trajectory selection. Furthermore, with virtual inputs, the lateral accelerations and the lead angle converge to the vicinity of zero at target interception. This results in maximization of the damage to the target due to high impact speed.
	\item Furthermore, unlike numerous existing guidance strategies \cite{cho2015nonsingular,Kumar2022BarLyap,SURVE2022406,Swati2023BarLyap, Liu2019IGCInpSat}, the proposed guidance scheme doesn't necessitate an ad-hoc way to restrict the interceptor's lateral acceleration to allowed bounds. The proposed approach provides a stronger guarantee of system stability by ensuring that the states never exceed their respective boundaries.
\end{itemize}

\section{Problem Formulation}
\begin{figure}[!ht]
	\centering
	\includegraphics[width = 0.3\textwidth]{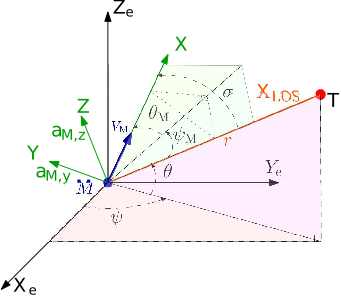}
	\caption{3D Interceptor-Target Engagement Geometry.}
	\label{fig:3D_engagement}
\end{figure}

Consider a three-dimensional engagement between an interceptor and a stationary target as shown in \Cref{fig:3D_engagement}. For the purpose of designing the guidance law, the autopilot is assumed to be sufficiently fast, and hence the interceptor can be considered to be a point-mass vehicle with constant speed. 
We consider three frames of reference: the inertial, body, and line-of-sight (LOS) frames, to describe the relative equations of motion between the interceptor and the target. The inertial frame is represented by a set of three mutually orthogonal axes, $\{X_{\rm e}, Y_{\rm e}, Z_{\rm e}\}$, while the body frame is denoted by another set of orthogonal axes, $(X, Y, Z)$.
The LOS frame is represented by $\{X_{\rm LOS}, Y_{\rm LOS}, Z_{\rm LOS}\}$. The angles $\psi$ and $\theta$ denote the azimuth and elevation angles of the LOS frame with respect to the inertial frame. The terms $\psi_\M$ and $\theta_\M$ represent the interceptor's velocity lead angles in the azimuth and elevation directions, measured from the LOS frame. Without loss of generality, let us assume that the interceptor is initially located at the origin of the inertial frame, with its velocity vector aligned with the $x$-axis of its body frame, denoted as $X$.
The kinematic equations, \cite{Song1994Kine3D}, describing the relative engagement between the interceptor and the target, in the spherical coordinate system, are given by
\begin{subequations}\label{eq:ENGKinematics}
	\begin{align}
		\dot{r} &= -V_\M \cos \theta_\M \cos \psi_\M, \\
		\dot{\theta} &= -\frac{V_\M \sin \theta_\M}{r}, \\
		\dot{\psi} &= -\dfrac{V_\M \cos \theta_\M \sin \psi_\M}{r \cos \theta}, \\
		\dot{\theta}_\M &= \dfrac{a_{\rm Mz}}{V_\M}-\dot{\psi} \sin \theta \sin \psi_\M-\dot{\theta} \cos \psi_\M,\label{eq:Ch4_3D_Kine_eqns_theta_M_dot} \\
		\nonumber\dot{\psi}_\M &= \dfrac{a_{\rm My}}{V_\M \cos \theta_\M}+\dot{\psi} \tan \theta_\M \cos \psi_\M \sin \theta-\dot{\psi} \cos \theta\\
		&\hspace{0.5cm} -\dot{\theta} \tan \theta_\M \sin \psi_\M.\label{eq:Ch4_3D_Kine_eqns_psi_M_dot}
	\end{align}
	\label{eq:Ch4_3D_Kine_eqns}
\end{subequations}
The interceptor lateral acceleration commands are denoted by $a_{\rm My}$ and $a_{\rm Mz}$, acting along the $Y$ and $Z$-axes of the body frame, respectively. Consequently, they directly influence the velocity lead angles along the azimuth ($\psi_\M$) and elevation ($\theta_\M$) directions, respectively. Throughout the rest of this article, $\theta_\M$ and $\psi_\M$ are also referred to as the heading angles for convenience. One can observe that the radial acceleration is not present in \eqref{eq:Ch4_3D_Kine_eqns}, which is reasonable to assume owing to the restricted throttle ability of solid rocket motors. Hence, the interceptor is directed solely through the lateral accelerations. During the terminal phase, the interceptor is assumed to be moving at a constant speed. It is evident from \eqref{eq:Ch4_3D_Kine_eqns} that the motions in the two lateral planes are highly coupled, and hence, decoupling the dynamics is to be avoided to prevent loss of performance in the case where the coupling is strong in nature. The effective lead angle, $\sigma$ and the heading angles, $\theta_\M$ and $\psi_\M$, are related by
\begin{equation}
	\cos{\sigma} = \cos{\psi_\M}~\cos{\theta_\M}.
	\label{eq:Ch4_sigma_psi_m_theta_m}
\end{equation}
Two essential temporal quantities in impact time-constrained guidance are defined in the following definition.  
\begin{definition}\cite{SRK2015largeHE}
	The time taken by the interceptor to intercept the target, starting from its launch instant or any pre-specified reference, is defined as the impact time. At any point during the engagement, the time-to-go refers to the remaining time before the target is intercepted. The desired value of time-to-go $t_{\rm go}^{\rm d}$ at any time instant $t_{\rm i}$, for a given desired impact time, $t_{\rm f}$, is given as $t_{\rm go}^{\rm d} = t_{\rm f} - t_{\rm i}$.
\end{definition}
\begin{assumption}\cite{zarchan2012tactical}\label{ch4_assum1}
	The angle of attack and the sideslip angles are assumed to be sufficiently small enough to be considered negligible. This, in turn, implies that the body frame coincides with the wind frame.
\end{assumption}
As a consequence of the above assumption, the seeker's FOV bound can be converted to a bound on the effective lead angle. Taking the FOV bound to be in the range $(-\pi/2, \pi/2)$, we get
\begin{equation}
	|\sigma| \le \sigma_{\max} < \pi/2,\label{Ch4_sigma_bound}
\end{equation}
where $\sigma_{\max}$ is the maximum permissible lead angle of the interceptor, which is one of the in-flight bounds considered in this work. Input saturation is treated as an additional in-flight constraint, and the complete problem statement is as below.
\begin{problem*}\label{prop_1}
	This paper aims to design a nonlinear guidance law that guarantees impact time-constrained interception while respecting the bounds on both the seeker's FOV bound as well as the interceptor's acceleration.
\end{problem*}
To exercise the Lyapunov theory of stability while proving the convergence of errors, the following results are presented.
\begin{lemma}
	\label{lem:Ch4_lyapunov}
	\cite{Khalil2002} Let $x=0$ be an equilibrium point for the system $\dot{x}=f(x)$. Let $V:\mathbb{R}^n\rightarrow \mathbb{R}$ be a continuously differentiable function such that
	\begin{equation*}
		V(0)=0\ \rm{and}\ V(x)>0\,\ \forall\ x\neq 0
	\end{equation*}
	\begin{equation*}
		||x||\rightarrow \infty \implies V(x) \rightarrow \infty
	\end{equation*}
	\begin{itemize}
		\item If $\dot{V}(x)<0\,\ \forall x\neq 0$, then $x=0$ is globally asymptotically stable.
		\item If $\dot{V}(x)<-kV^\rho(x)\,\ \forall x\neq 0$, where $k>0$ and $\rho \in (0,1)$, then $x=0$ is finite time stable.
	\end{itemize}
\end{lemma}
We aim to design guidance commands within a nonlinear framework to prevent any possible degradation in performance that may arise in engagement scenarios with large initial deviations. 

\section{Main results}
In real-world applications, interceptors are subject to control limitations and actuator-induced lag when applying control inputs. These are the major features of autopilots that affect the design of the guidance strategy. Thus, inspired by \cite{Zou2017InpSat}, we propose a smooth input saturation model to incorporate the bounded control availability and lag as 
\begin{equation}\label{eq:Zou_a_dot}
	\dot{a}_{\rm M} = \left[ 1-\left({a_{\rm M}}/{a_{\max}}\right)^n \right]b - \rho a_{\rm M},
\end{equation}
where $n\ge 2$ and even, and $\rho>0$ is a small real number. In \eqref{eq:Zou_a_dot}, the variable $a_{\rm {M}}$ is the control input (lateral acceleration) that acts on the interceptor, and $b$ is an auxiliary control input (commanded acceleration). The inclusion of the term $-\rho a_{\rm M}$ causes $\dot{a}_\M$ to reach zero prior to $a_{\rm M}$ reaching $a_{\max}$. This results in $\rho$ providing a safety margin between the maximum achieved acceleration and $a_{\max}$. 
\subsection{Time Constrained Guidance with Bounded FOV and Input}\label{Sec3B:3D}
In this section, we derive the necessary guidance commands
to achieve the objectives of \Cref{prop_1} in a three-dimensional setting whose engagement kinematics are given in \eqref{eq:ENGKinematics}. To proceed, we define the range and lead angle errors analogous to those in \cite{H_J_Kim} as follows:
\begin{subequations}\label{eq:ch4_e1e2}
	\begin{align}
		z_1 &= V_\M t_{\rm go}^{\rm d} - r,\label{eq:ch4_e1}\\
		z_2 &= \sigma - \sigma_{\rm d}, \dot{z}_2 = \dot{\sigma}-\dot{\sigma}_{\rm d} \label{eq:ch4_z2}
	\end{align}
\end{subequations}
where the term $\sigma_{\rm d}$ is a virtual input that can be selected to ensure $z_1$ converges to zero. In \cite{H_J_Kim}, the authors have shown that the choice of the desired lead angle as
\begin{equation}\label{eq:Ch4_sigmad_kim}
	\sigma_{\rm d} = \cos^{-1}(1-k_1\sgmf(z_1)),
\end{equation}
where $\sgmf(\cdot)$ is an approximated signum function given by
\begin{equation}
	{\sgmf}(x) = 
	\begin{cases}
		-\dfrac{x^3}{2\phi^3}+\dfrac{3x}{2\phi},&|x|\le\phi\\
		\sign(x),& |x|>\phi
	\end{cases},
	\label{sgmf}
\end{equation}
ensures linear convergence of the range error $z_1$ for $|z_1(t)|>\phi$ and asymptotic convergence for $|z_1(t)|<\phi$. One can also vary the parameter $k_1$ to control the maximum attainable lead angle throughout the engagement.
\begin{remark}\label{rem:Ch4_rem_2}
	One should note that by substituting the lead angle bound, $\sigma_{\max}$, into the $\sigma_{\rm d}$ as given in \eqref{eq:Ch4_sigmad_kim}, the upper bound on the value of $k_1$ can be obtained as $k_1 < 1 - \cos{\sigma_{\max}}$.
\end{remark}
Similar to the expression of lead angle error, $z_2$ in \eqref{eq:ch4_e1e2}, we define heading angle errors, $z_3$ and $z_4$, as
\begin{equation}\label{eq:T10_z2z3}
	z_3 = \theta_\M - \theta_{\rm M_d}, \quad 
	z_4 = \psi_\M - \psi_{\rm M_d}, 
\end{equation}
where the terms $\theta_{\rm M_d}$ and $\psi_{\rm M_d}$ are the desired heading angles. Following the approach in \cite{AshokIFAC}, the following choice of $\theta_{\rm M_d}$ and $\psi_{\rm M_d}$ ensures the convergence of the range error, provided that the heading angle errors converge.
\begin{equation}\label{eq:T10_bk_step_psi_theta_m_2}
	\psi_{\rm M_d} = \theta_{\rm M_d} = \dfrac{1}{2}\cos^{-1}\left[2\cos(\sigma_{\rm d})-1\right].
\end{equation}
where $\sigma_{\rm d}$ is as given in
\begin{equation}\label{eq:T10_sigma_d_kim}
	\sigma_{\rm d} = f(z_1) = \cos^{-1}[1-k_1{\sgmf}(z_1)].
\end{equation}
One can refer to \cite{AshokIFAC} for a detailed proof.
In the case of 3D engagement scenarios, one may utilize the acceleration saturation model given in \eqref{eq:Zou_a_dot} to obtain the lateral accelerations in both $Y$ and $Z$ directions follows
\begin{subequations}\label{eq:T10_ay_dot_az_dot}
	\begin{align}
		\dot{a}_{\rm My} = \left[ 1-\left(\dfrac{a_{\rm My}}{a_{\rm y,\max}}\right)^n \right] b_{\rm y} - \rho a_{\rm My},
		\label{eq:T10_ay_dot}\\
		\dot{a}_{\rm Mz} = \left[ 1-\left(\dfrac{a_{\rm Mz}}{a_{\rm z,\max}}\right)^n \right] b_{\rm z} - \rho a_{\rm Mz},
		\label{eq:T10_az_dot}
	\end{align}
\end{subequations}
where $n\ge 2$ and even, $\rho>0$ and small, while the terms $b_{\rm y}$ and $b_{\rm z}$ are the commanded inputs in respective directions. Next, we define acceleration errors $z_{y}$ and $z_{z}$ as 
\begin{equation}\label{eq:T10_zyzz}
	z_{\rm y} = a_{\rm My} - \alpha_{\rm y},~~ 
	z_{\rm z} = a_{\rm Mz} - \alpha_{\rm z}, 
\end{equation} 
with $\alpha_{\rm y}$ and $\alpha_{\rm z}$ are the stabilizing functions. Therefore, our aim is to design $(b_{\rm y}, b_{\rm z})$ that will ensure the errors, $(z_3, z_{\rm z})$ and $(z_4, z_{\rm y})$, converge to zero. Essentially, we aim to formulate a guidance strategy that achieves interception at a desired impact time while adhering to input and FOV constraints in a 3D setting. Owing to this, we present the following theorem, which will give the commanded input for the 3D engagements.
\begin{theorem}
	\label{thm:ACC_thm3}
	For the engagement kinematics given in \eqref{eq:Ch4_3D_Kine_eqns} and the input saturation models as in \eqref{eq:T10_ay_dot_az_dot}, if the commanded inputs are designed as 
	\begin{subequations}\label{eq:T10_bybz_1}
		\begin{align}
			b_{\rm y} =& \frac{\rho a_{\rm My} + \dot{\alpha}_{\rm y} - z_4/(V_\M \cos\theta_\M ) - k_{\rm y}z_{\rm y}}{\left[1-\left(a_{\rm My}/a_{\rm y,\max}\right)^n \right]}, \label{eq:T10_by_1}\\
			b_{\rm z} =& \dfrac{\rho a_{\rm Mz} + \dot{\alpha}_{\rm z} - z_3/V_\M - k_{\rm z}z_{\rm z}}{\left[ 1-\left({a_{\rm Mz}}/{a_{\rm z,\max}}\right)^n \right]},\label{eq:T10_bz_1}
		\end{align}
	\end{subequations}
	with the terms $\dot{\alpha}_y$ and $\dot{\alpha}_z$ as
	\begin{subequations}\label{eq:T10_ayc_dot_azc_dot_1}
		\begin{align}
			\nonumber \dot{\alpha}_{\rm y} &= -V_{\rm{M}} \sin{\theta_{\rm M}}\dot{\theta}_{\rm M}\left[-\dot{\psi}\tan{\theta_{\rm M}}\cos{\psi_{\rm M}}\sin{\theta} + \dot{\psi}\cos{\theta} \right.\\
			\nonumber &\left.+ \dot{\theta}\tan{\theta_{\rm M}}\sin{\psi_{\rm M}} + \dot{\psi}_{\rm M_d} - k_4 z_4\right]\\
			\nonumber& + V_{\rm M} \cos{\theta_{\rm M}} \left[ -\ddot{\psi}\tan{\theta_{\rm M}}\cos{\psi_{\rm M}}\sin{\theta}\right. \\
			\nonumber &\left.-\dot{\psi}\dot{\theta}_\M\sec^2{\theta_{\rm M}}\cos{\psi_{\rm M}}\sin{\theta} +\dot{\psi}\dot{\psi}_{\rm M}\tan{\theta_{\rm M}}\sin{\psi_{\rm M}}\sin{\theta} \right.\\
			\nonumber& \left.-\dot{\psi}\dot{\theta}\tan{\theta_{\rm M}}\cos{\psi_{\rm M}}\cos{\theta} - \dot{\psi}\dot{\theta}\sin{\theta} + \ddot{\psi}\cos{\theta}\right.\\
			\nonumber & \left. + \ddot{\theta}\tan{\theta_{\rm M}}\sin{\psi_{\rm M}} + \dot{\theta}\dot{\theta}_{\rm M}\sec^2{\theta_{\rm M}}\sin{\psi_{\rm M}}\right.\\ 
			&\left. +\dot{\theta}\dot{\psi}_{\rm M}\tan{\theta_{\rm M}}\cos{\psi_{\rm M}}
			+ \ddot{\psi}_{\rm M_d} -k_4\,\dot{z}_4\right],\label{eq:T10_ayc_dot_1}\\
			\nonumber \dot{\alpha}_{\rm z} &= V_\M \left[ \ddot{\psi}\sin{\theta}\sin{\psi_{\rm M}} + \dot{\psi}\dot{\theta}\cos{\theta}\sin{\psi_{\rm M}} + \dot{\psi}\dot{\psi}_{\rm M}\sin{\theta}\right.\\
			&\left.\cos{\psi_{\rm M}}+ \ddot{\theta}\cos{\psi_{\rm M}} - \dot{\theta}\dot{\psi}_{\rm M}\sin{\psi_{\rm M}} + \ddot{\theta}_{\rm M_d} -k_3\dot{z}_3\right],\label{eq:T10_azc_dot_1}
		\end{align}
	\end{subequations}
	where $k_3$, $k_4$, $k_{\rm y}$, and $k_{\rm z}$ are positive constants, then the errors, $(z_3, z_{\rm z})$ and $(z_4, z_{\rm y})$, will converge to zero asymptotically and a successful target interception will be achieved at the desired impact time while adhering the FOV and input constraints.
\end{theorem}
\begin{proof}
	Consider the Lyapunov function candidates, $\mathcal{V}_1$, and $\mathcal{V}_2$, where $\mathcal{V}_1 = 0.5{{z_3}^2}$, and $\mathcal{V}_2 = 0.5{{z_4}^2}$, respectively. On differentiating, $\mathcal{V}_1$, and $\mathcal{V}_2$ with respect to time and substituting for $\dot{\theta}_\M $ and $\dot{\psi}_\M $ from \eqref{eq:Ch4_3D_Kine_eqns}, we get
	\begin{subequations}\label{eq:T10_LyaFuncDot_341}
		\begin{align}
			\nonumber\dot{\mathcal{V}}_1 &= {z_3} \left( \dfrac{a_{\rm Mz}}{V_\M }-\dot{\psi} \sin \theta \sin \psi_\M -\dot{\theta} \cos \psi_\M - \dot{\theta}_{\rm M_d} \right),\\ 
			\nonumber\dot{\mathcal{V}}_2 &= {z_4} \left( \dfrac{a_{\rm My}}{V_\M \cos \theta_\M }+\dot{\psi} \tan \theta_\M \cos \psi_\M \sin \theta-\dot{\psi} \cos \theta \right. \\
			\nonumber&\hspace{0.5cm} \left. -\dot{\theta} \tan \theta_\M \sin \psi_\M - \dot{\psi}_{\rm M_d} \right).
		\end{align}
	\end{subequations}
	On substituting $a_{\rm My}$ and $a_{\rm Mz}$ from \eqref{eq:T10_zyzz}, one may obtain
	\begin{align}
		\nonumber \dot{\mathcal{V}}_1 &= {z_3} \left( \dfrac{z_z+\alpha_{\rm z}}{V_\M }-\dot{\psi} \sin \theta \sin \psi_\M -\dot{\theta} \cos \psi_\M - \dot{\theta}_{\rm M_d} \right),\\ 
		\dot{\mathcal{V}}_2 &= {z_4} \left(\dfrac{z_{\rm y}+\alpha_{\rm y}}{V_\M \cos \theta_\M }+\dot{\psi} \tan \theta_\M \cos \psi_\M \sin \theta \label{eq:T10_LyaFuncDot_42} \right. \\
		\nonumber &\hspace{0.5cm} \left.-\dot{\psi} \cos \theta -\dot{\theta} \tan \theta_\M \sin \psi_\M - \dot{\psi}_{\rm M_d} \right). 
	\end{align}
	On choosing stabilizing functions, $\alpha_{\rm y}$ and $\alpha_{\rm z}$ as
	\begin{subequations}\label{eq:T10_ayc_azc_1}
		\begin{align}
			\nonumber \alpha_{\rm y} &= V_\M \cos{\theta_\M } \left[ -\dot{\psi}\tan{\theta_\M }\cos{\psi_\M }\sin{\theta}+ \dot{\psi}\cos{\theta}\right.\\
			&\left. + \dot{\theta}\tan{\theta_\M }\sin{\psi_\M } + \dot{\psi}_{\rm M_d} -k_4\,z_4\right],\\
			\alpha_{\rm z} &= V_\M \left[ \dot{\psi}\sin{\theta}\sin{\psi_\M } + \dot{\theta}\cos{\psi_\M } + \dot{\theta}_{\rm M_d} -k_3\,z_3\right],
		\end{align}
	\end{subequations}
	and substituting them in \eqref{eq:T10_LyaFuncDot_42}, would result into
	\begin{align}\label{eq:T10_LyaFuncDot_343}
		\dot{\mathcal{V}}_1 &= -k_3\,{z_3}^2 + \dfrac{z_3\,z_{\rm z}}{V_\M },
		\dot{\mathcal{V}}_2 = -k_4\,{z_4}^2 + \dfrac{{z_4}\,z_{\rm y}}{V_\M \cos \theta_\M }.
	\end{align}
	Next to account for the input constraints, consider the augmented Lyapunov function candidates, $\mathcal{V}_{z}$ and $\mathcal{V}_{y}$, the expression for which is given as $\mathcal{V}_{z} = 0.5{z_3}^2 + 0.5{z_{\rm z}}^2 = \mathcal{V}_1 + 0.5{z_{\rm z}}^2$, and $\mathcal{V}_{y} = 0.5{z_4}^2 + 0.5{z_{\rm y}}^2 = \mathcal{V}_{2} + 0.5{z_{\rm y}}^2$.
	On taking time differentiation of $\mathcal{V}_{z}$ and $\mathcal{V}_{y}$, and substituting \eqref{eq:T10_LyaFuncDot_343} and \eqref{eq:T10_ay_dot_az_dot}, we get
	\begin{align}\label{eq:T10_LyaFuncDot_3z4y2}
		\nonumber \dot{\mathcal{V}}_{z} &= -k_3\,{z_3}^2 + z_{\rm z} \left\{ \dfrac{z_3}{V_\M } + \left[ 1-\left({a_{\rm Mz}}/{a_{\rm z,\max}}\right)^n \right] b_{\rm z}\right.\\
		\nonumber &\left.- \rho a_{\rm Mz} - \dot{\alpha}_{\rm z}\right\},\\ 
		\nonumber\dot{\mathcal{V}}_{y} &= -k_4\,{z_4}^2 + z_{\rm y}\left\{ \left({z_4}/{V_\M \cos \theta_\M}\right) - \rho a_{\rm My} - \dot{\alpha}_{\rm y}\right. \\
		&\left. + \left[ 1-\left({a_{\rm My}}/{a_{\rm y,\max}}\right)^n \right] b_{\rm y} \right\}.
	\end{align} 
	Choosing the commanded inputs as in \eqref{eq:T10_bybz_1}, and substituting them in \eqref{eq:T10_LyaFuncDot_3z4y2}, one may obtain 
	\begin{align}\label{eq:T10_LyaFuncDot_3z4y3}
		\dot{\mathcal{V}}_{z} &= -k_3\,z^2_3 - k_{\rm z}\,z^2_{\rm z},~~ 
		\dot{\mathcal{V}}_{y} = -k_4\,z^2_4 - k_{\rm y}\,z^2_{\rm y},  
	\end{align} 
	which may further be simplified to $\dot{\mathcal{V}}_{z} = -k_{3\rm z}\mathcal{V}_{z}$, and $\dot{\mathcal{V}}_{y} = -k_{4\rm y}\mathcal{V}_{y}$,
	where $k_{3\rm z} = \min\{k_3,k_{\rm z}\}$ and $k_{4\rm y} = \min\{k_4,k_{\rm y}\}$. This confirms asymptotic convergence of errors, $z_3$, $z_{\rm z}$, $z_4$, and $z_{\rm y}$, by \Cref{lem:Ch4_lyapunov}. Therefore, a successful target interception will be achieved at the desired impact time while obeying the FOV and input constraints of the interceptor. This completes the proof. 
\end{proof}
\begin{remark}\label{rem:rem_3D_theorem}
	The terms $\dot{\alpha}_{\rm y}$ and $\dot{\alpha}_{\rm z}$ in \eqref{eq:T10_ayc_dot_azc_dot_1} consisted of the terms $\ddot{\theta}$, $\ddot{\psi}$, $\ddot{\theta}_{\rm M_d}$, $\ddot{\psi}_{\rm M_d}$, $\dot{z}_3$ and $\dot{z}_4$, which can be obtained as follows: 
	\begin{subequations}\label{eq:t10_alpy_alpz_Dot_supportEqns}
		\begin{align}
			\ddot{\theta} &= \dfrac{V_\M \sin{\theta_\M}}{r^2}\dot{r} - \dfrac{V_\M \cos{\theta_\M}}{r}\dot{\theta}_\M\\
			\nonumber \ddot{\psi} &= \dfrac{V_\M \cos{\theta_\M} \sin{\psi_\M} }{r^2 \cos{\theta}}\dot{r} -\dfrac{V_\M \cos{\theta_\M} \sin{\psi_\M} \sin{\theta}}{r \cos^2{\theta}}\dot{\theta} \\
			& -\dfrac{V_\M \cos{\theta_\M} \cos{\psi_\M} }{r \cos{\theta}}\dot{\psi}_\M+\dfrac{V_\M \sin{\theta_\M} \sin{\psi_\M} }{r \cos{\theta}}\dot{\theta}_\M\\
			\dot{z}_3 &= \dot{\theta}_\M - \dot{\theta}_{\rm M_d}, \dot{z}_4 = \dot{\psi}_\M - \dot{\psi}_{\rm M_d}\\ 
			\dot{\psi}_{\rm M_d} &= \dfrac{\sin{\sigma_{\rm d}}}{\sin(2\psi_{\rm M_d})}\dot{\sigma}_{\rm d}, \dot{\theta}_{\rm M_d} = \dfrac{\sin{\sigma_{\rm d}}}{\sin(2\theta_{\rm M_d})}\dot{\sigma}_{\rm d},\\
			\ddot{\psi}_{\rm M_d} &= \dfrac{\ddot{\sigma}_{\rm d} \sin{\sigma_{\rm d}} + (\dot{\sigma}_{\rm d})^2 \cos{\sigma_{\rm d}} - 2(\dot{\psi}_{\rm M_d})^2 \cos{2\psi_{\rm M_d}} }{\sin(2\psi_{\rm M_d})},\\
			\ddot{\theta}_{\rm M_d} &= \dfrac{\ddot{\sigma}_{\rm d} \sin{\sigma_{\rm d}} + (\dot{\sigma}_{\rm d})^2 \cos{\sigma_{\rm d}} - 2(\dot{\theta}_{\rm M_d})^2 \cos{2\theta_{\rm M_d}} }{\sin(2\theta_{\rm M_d})},\\
			\dot{\sigma}_{\rm d} =& \dfrac{k_1}{\sin{\sigma_{\rm d}}}\left[\dfrac{d}{dz_1}\sgmf(z_1)\right] \dot{z}_1,
			\label{eq:T10_sigmad_Dot_2}\\
			\nonumber\ddot{\sigma}_{\rm d} =& \dfrac{1}{\sin{\sigma_{\rm d}}} \left\{ k_1\left[\dfrac{d^2}{(dz_1)^2}\sgmf(z_1)\right] (\dot{z}_1)^2\right.\\
			& \left. + k_1\left[\dfrac{d}{dz_1}\sgmf(z_1)\right] \ddot{z}_1 -(\dot{\sigma}_{\rm d})^2\cos{\sigma_{\rm d}} \right\}.
			\label{eq:T10_sigmad_DDot_2}
		\end{align}
	\end{subequations}
	With the derivatives defined as
	\begin{align*}
		\dfrac{d}{dz_1}\sgmf(z_1) &= 
		\begin{cases}
			\dfrac{-3z^2_1}{2\phi^3} + \dfrac{3}{2\phi}&, \mathrm{if}~~|z_1|\le\phi\\
			0&, \mathrm{else}
		\end{cases},\\
		\dfrac{d^2}{(dz_1)^2}\sgmf(z_1) &= 
		\begin{cases}
			\dfrac{-3z_1}{\phi^3}&, \mathrm{if}~~|z_1|\le\phi\\
			0&, \mathrm{else}
		\end{cases}.
	\end{align*}
	Equations in \eqref{eq:T10_ayc_dot_azc_dot_1} and  \eqref{eq:t10_alpy_alpz_Dot_supportEqns}, complete the guidance law proposed in \eqref{eq:T10_bybz_1}.
\end{remark}

\begin{corollary}
	With the proposed guidance strategy, both the velocity lead angle, $\sigma$, and the lateral acceleration components, $a_{\rm {My}}$ and $a_{\rm {Mz}}$, converge in the neighborhood of zero near target interception. 
\end{corollary}
\begin{proof}
	As time $t\rightarrow t_{\rm{f}}$, the error $z_1 = 0$, which in turn implies $\sigma_{\rm d} = 0$, from \eqref{eq:ch4_z2}. Once, $\sigma_{\rm d} = 0$, it follows from \eqref{eq:T10_bk_step_psi_theta_m_2}, that the terms $\theta_{\rm d} = \psi_{\rm d} = 0$. From \eqref{eq:T10_LyaFuncDot_3z4y3}, we have 
	\begin{align}\label{eq:Corollary_theta_psi}
		\dot{\mathcal{V}}_{z} &= -k_3\,z^2_3 - k_{\rm z}\,z^2_{\rm z},~~ 
		\dot{\mathcal{V}}_{y} = -k_4\,z^2_4 - k_{\rm y}\,z^2_{\rm y},  
	\end{align} 
	which implies asymptotic convergence of errors, ${z_3}$, $z_{\rm z}$, ${z_4}$, and $z_{\rm y}$. Once ${z_3} = z_{\rm z} = {z_4}$ = ${z_{\rm y}} = 0$, we have $\theta_{\rm{M_d}} = \psi_{\rm{M_d}} = \theta_{\rm M} = \psi_{\rm M}$, implying that the heading angles converge to zero. Furthermore, with $z_{\rm z} = z_{\rm y} = 0$, we obtain, $a_{\rm{My}} = \alpha_{\rm y} $ and $a_{\rm{Mz}} = \alpha_{\rm z}$. Consequently, from \ref{eq:T10_ayc_azc_1}, it follows that $\alpha_{\rm y} = \alpha_{\rm z} = 0$ (as $\psi_{\rm M} = \theta_{\rm M} = 0$). Hence, the lateral acceleration components satisfy $a_{\rm{My}} = a_{\rm{Mz}} = 0$. This completes the proof.
\end{proof}

\subsection{Special Case of Planar Engagement Scenarios} \label{Sec:Inp_Sat_2D}
In this section, we consider the two-dimensional (planar) engagement setting. The kinematic equations can be obtained either by taking $\psi_{\rm M} = \psi = 0$ or by setting $\theta_{\rm M} = \theta = 0$. Without loss off generality, let us assume $\theta = \theta_\M = a_{\rm Mz} =0$, which further yields $\dot{\theta} = \dot{\theta}_\M=0$. Thus, the corresponding kinematic equations are as follows:
\begin{align}
	\dot{r} = -V_\M \cos{\sigma},~
	\dot{\theta} = -\frac{V_\M \sin{\sigma}}{r},~
	\dot{\sigma} = \dfrac{a_{\rm My}}{V_\M} - \dot{\theta}.
	\label{eq:Ch4_2D_Kine_eqns}     
\end{align}
To ensure that the interceptor's lateral acceleration remains within allowable limits, we employ the acceleration saturation model given in \eqref{eq:Zou_a_dot}. By considering the lateral acceleration, $a_{\rm My}$, as an additional state, the system can be reformulated such that  $b_{\rm y}$ is the control input to be designed. Hence, we append \eqref{eq:Zou_a_dot} to the kinematic equations of the interceptor and proceed to design the guidance command, which would be an expression for $b_{\rm y}$.
Next, to design a guidance command, we define the following
\begin{equation}\label{eq:ch4_e3}
	z_y = a_{\rm My} - \alpha_{\rm y},
\end{equation}
where $\alpha_y$ is a stabilizing function. The following theorem presents the commanded input, $b_{\rm y}$, that ensures asymptotic convergence of the errors.
\begin{theorem}
	\label{thm:ACC_thm1} 
	Consider the engagement kinematics given in \eqref{eq:Ch4_2D_Kine_eqns} with the input saturation model as in \eqref{eq:Zou_a_dot}. If the commanded input, $b_{\rm y}$ is chosen as 
	\begin{equation}
		b_{\rm y} = \dfrac{\rho a_{\rm My} + \dot{\alpha}_{\rm y} - (z_2/V_\M) - k_yz_y}{\left[ 1-\left(\dfrac{a_{\rm My}}{a_{\rm y,max}}\right)^n \right]},
		\label{eq:T4_b_1}
	\end{equation}
	where the term $\dot{\alpha}_{\rm y}$ is given by
	\begin{align}\label{eq:T4_ac_dot}
		\dot{\alpha}_{\rm y} & = V_\M \left[ \ddot{\sigma}_{\rm d} - \dfrac{V_\M \cos{\sigma}}{r}\dot{\sigma} + \dfrac{V_\M \sin{\sigma}}{r^2} \dot{r} - k_2 \left( \dot{\sigma} - \dot{\sigma}_{\rm d} \right)\right],
	\end{align}
	with $n\geq 2$ (even), $\rho > 0$ (a small real number), and $k_y$, $k_2$ as positive constants,
	then both the interceptor's lead angle error, $z_2$, and acceleration error, $z_y$, will converge to zero asymptotically. Consequently, the interceptor will successfully intercept the target at a desired impact time, maintaining the lead angle and input within their respective bounds.
\end{theorem}
\begin{proof}\label{ProofTHM1}
	Consider the Lyapunov function candidate given by $\mathcal{V}_3 = 0.5{z^2_2}$. On differentiating  $\mathcal{V}_3$ with respect to time and substituting $\dot{z}_2$ and $\dot{\sigma}$ from \label{eq:ch4_z2} and \eqref{eq:Ch4_2D_Kine_eqns}, respectively, we get
	\begin{align}
		\nonumber\dot{\mathcal{V}}_3 &= z_2\,\dot{z}_2 = z_2 \left( \dot{\sigma} - \dot{\sigma}_{\rm d} \right) = z_2 \left( \dfrac{a_{\rm My}}{V_\M} - \dot{\theta} - \dot{\sigma}_{\rm d} \right)\\
		&= z_2 \left( \dfrac{a_{\rm My}}{V_\M} + \dfrac{V_\M \sin{\sigma}}{r} - \dot{\sigma}_{\rm d} \right).
		\label{eq:T4_LyaFuncDot_11}
	\end{align}
	Substituting $a_{\rm My}$ from \eqref{eq:ch4_e3}, \eqref{eq:T4_LyaFuncDot_11} reduces to
	\begin{equation}\label{eq:T4_LyaFuncDot_12}
		\dot{\mathcal{V}}_3 = z_2 \left( \dfrac{z_y + \alpha_{\rm y}}{V_\M} + \dfrac{V_\M \sin{\sigma}}{r} - \dot{\sigma}_{\rm d} \right).
	\end{equation}
	On choosing stabilizing function, $\alpha_{\rm y}$ as $\alpha_{\rm y} = V_\M \left[ \dot{\sigma}_{\rm d} - \left(V_\M (\sin{\sigma})/{r}\right) -k_2z_2 \right]$, and substituting it in $\dot{\mathcal{V}}_3$, results in $\dot{\mathcal{V}}_3 = -k_2z_2^2 + \frac{z_2z_y}{V_\M}$.
	To incorporate the input constraint, we consider another Lyapunov function candidate $\mathcal{V}$, as $\mathcal{V} = \mathcal{V}_3 + \mathcal{V}_4$, where $\mathcal{V}_4 = 0.5z^2_{\rm y}$. On differentiating $\mathcal{V}$ with respect to time and substituting \eqref{eq:Zou_a_dot} and $\dot{\mathcal{V}}_3$, one may obtain
	\begin{align}\label{eq:T4_LyaFuncDot_21}
		\nonumber \dot{\mathcal{V}} &= \dot{\mathcal{V}}_3 + \dot{\mathcal{V}}_4 = -k_2z_2^2 + \dfrac{z_2z_y}{V_\M} + z_y\,\dot{z}_{\rm y}\\
		\nonumber &= -k_2z_2^2 + z_y\left\{ \dfrac{z_2}{V_\M} + \left[ 1-\left(\dfrac{a_{\rm My}}{a_{\rm y,max}}\right)^n \right] b_{\rm y} \right.\\
		&\left.- \rho a_{\rm My} - \dot{\alpha}_{\rm y} \right\}.
	\end{align}
	On choosing $b_{\rm y}$ as in \eqref{eq:T4_b_1}, \eqref{eq:T4_LyaFuncDot_21} simplifies to $\dot{\mathcal{V}} = -k_2 z_2^2 - k_y z_y^2$.
	By selecting $k_{\rm 2y} = \min\{k_2,k_y\}$, where $k_2$ and $k_y$ are positive constants, one may write $\dot{\mathcal{V}} = -2 k_{\rm 2y}\mathcal{V}$, which further implies that $\dot{\mathcal{V}}$ is negative definite. This confirms the asymptotic convergence of the error $z_2$ and $z_y$, by \Cref{lem:Ch4_lyapunov}.  
\end{proof}
The expression of the term $\dot{\alpha}_y$ in \eqref{eq:T4_ac_dot} requires the computation of, $\dot{\sigma_{\rm d}}$, and $\ddot{\sigma}_{\rm d}$, which can be obtained from \eqref{eq:T4_ac_dot}, and \Cref{rem:rem_3D_theorem}, respectively.
\begin{remark}
	Note that the commanded input in \eqref{eq:T4_b_1} depends on $\ddot{\sigma}_{\rm d}$. However, the expression for $\sigma_{\rm d}$ in \eqref{eq:Ch4_sigmad_kim} is only $C^1$ continuous, which implies $b_{\rm y}$ is discontinuous at $|z_1|=\phi$. According to the input saturation model in \eqref{eq:Zou_a_dot}, $b_{\rm y}$ affects the rate of change of acceleration and not the instantaneous acceleration. Hence, there is no requirement for the commanded input to be continuous.
\end{remark}
\section{Performance study}
The performance of the guidance laws proposed in \eqref{eq:T10_bybz_1} and \eqref{eq:T4_b_1} is evaluated through numerical simulations in this section. 

Unless stated otherwise, the simulation parameters for all upcoming results are listed in \Cref{tab:T10_3_Sim_params}. The parameter $k_1$ is chosen according to the requirement specified in  \Cref{rem:Ch4_rem_2}. Additionally, the initial heading is defined by the heading angles, $\theta_\m(0)$ and $\psi_\m(0)$.
\renewcommand{\arraystretch}{1}
\setlength{\tabcolsep}{0pt}
\begin{table}[H]
	\centering
	\caption{Simulation parameters.}
	\label{tab:T10_3_Sim_params}
	\begin{tabular}{lcc}
		\hline\hline
		Parameter & Value & Unit\\ \hline
		Interceptor speed ($V_\m $) & $250$ & $\rm ms^{-1}$\\ 
		($x(0),y(0),z(0)$) & $(-10,0,0)$ & $\rm km$\\
		Impact time ($t_{\rm f}$) & $50$ & $\rm s$\\
		Launch angle $(\theta_\m (0), \psi_\m (0))$ & $(-10,10)$ & $^\circ$\\
		Target coordinates ($x_{\rm T},y_{\rm T},z_{\rm T}$) & $(0,0,0)$ & $\rm km$\\
		$N$ & 3 & $-$\\
		$k_1$ & $1-\cos{\sigma_{\max}} -0.01$ & $-$\\
		$k_3$ & 1 & $-$\\
		$k_4$ & 1 & $-$\\
		$k_y$ & 7 & $-$\\
		$k_z$ & 7 & $-$\\
		$\phi$ & 300 & $\rm m$\\
		$\rho$ & 0.1 & $-$\\
		$n$ & 2 & $-$\\
		Acceleration magnitude limit ($a_{\rm max}$) & $10$g & $\rm ms^{-2}$\\
		Seeker's FOV bound ($\sigma_{\rm max}$) & $60$ & $^\circ$\\
		\hline\hline
	\end{tabular}
\end{table}
\subsection{Three-dimensional engagements with different impact time and initial headings}
\begin{figure*}[!h]
	\centering
	\begin{subfigure}[b]{0.245\linewidth}
		\includegraphics[width=\linewidth]{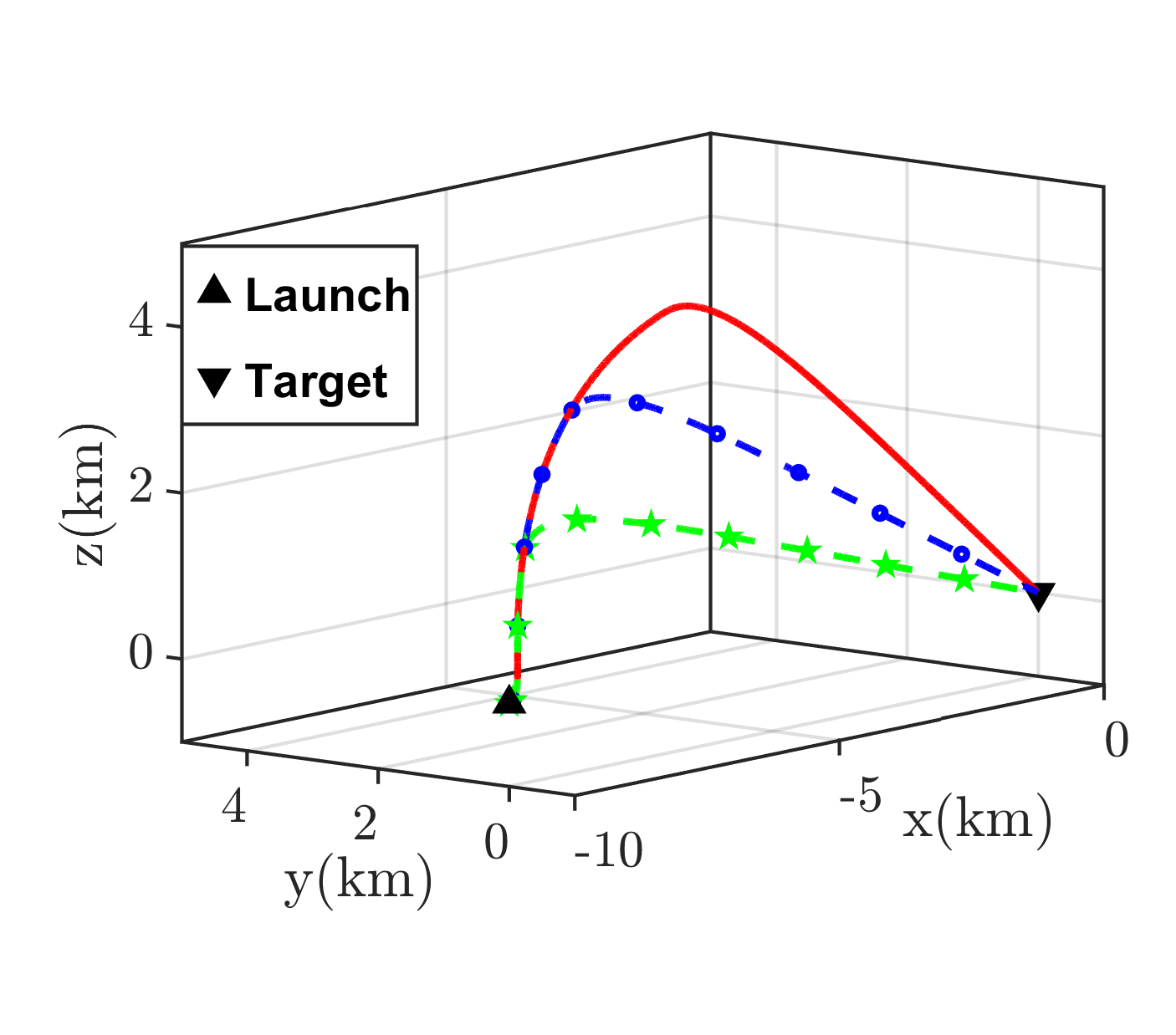}
		\caption{Trajectory.}\label{fig:T10_diff_tf_trajectory}
	\end{subfigure}
	\begin{subfigure}[b]{0.245\linewidth}
		\includegraphics[width=\linewidth]{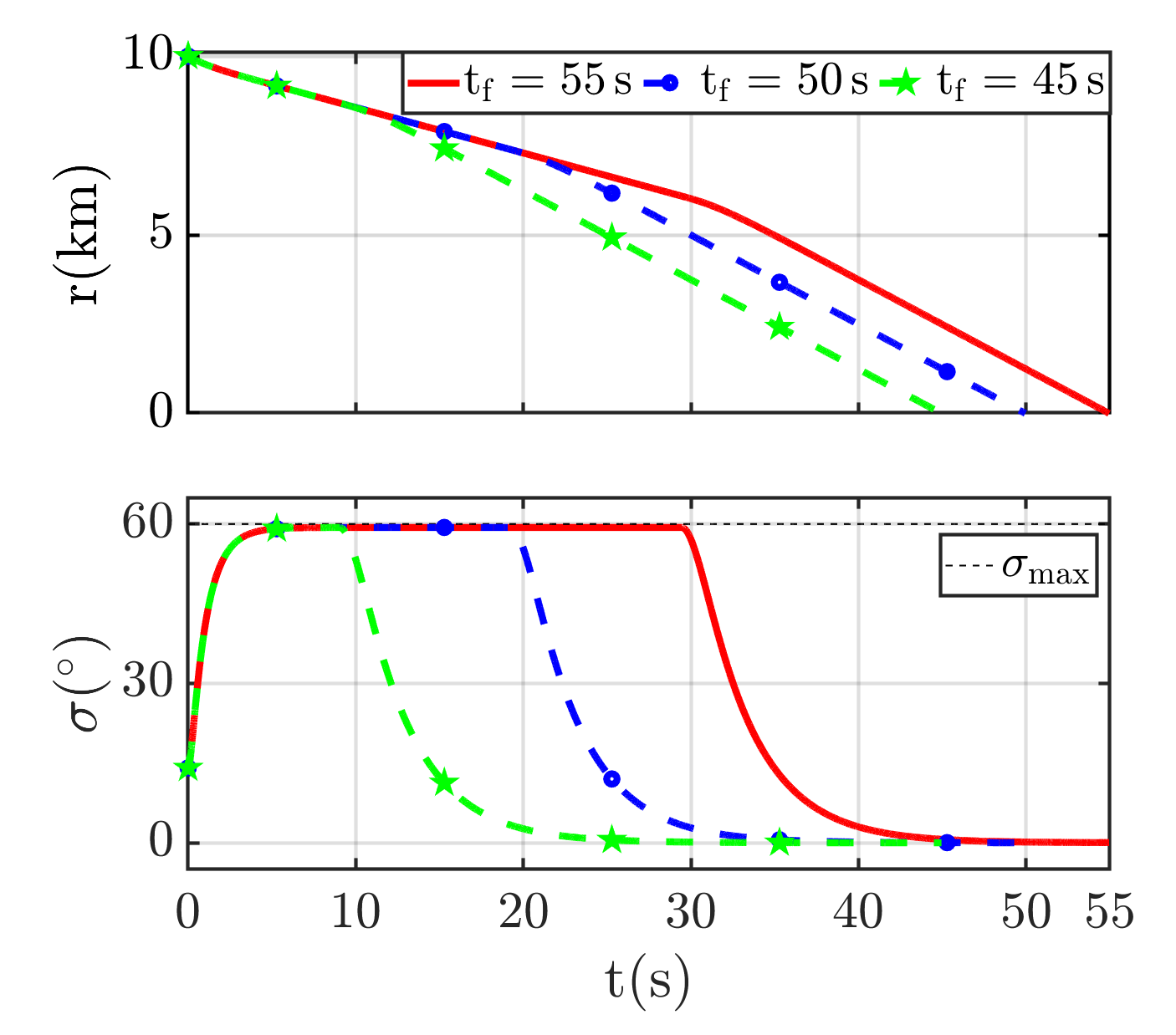}
		\caption{Range and lead angle.}\label{fig:T10_diff_tf_r_sigma}
	\end{subfigure}
	\begin{subfigure}[b]{0.245\linewidth}
		\includegraphics[width=\linewidth]{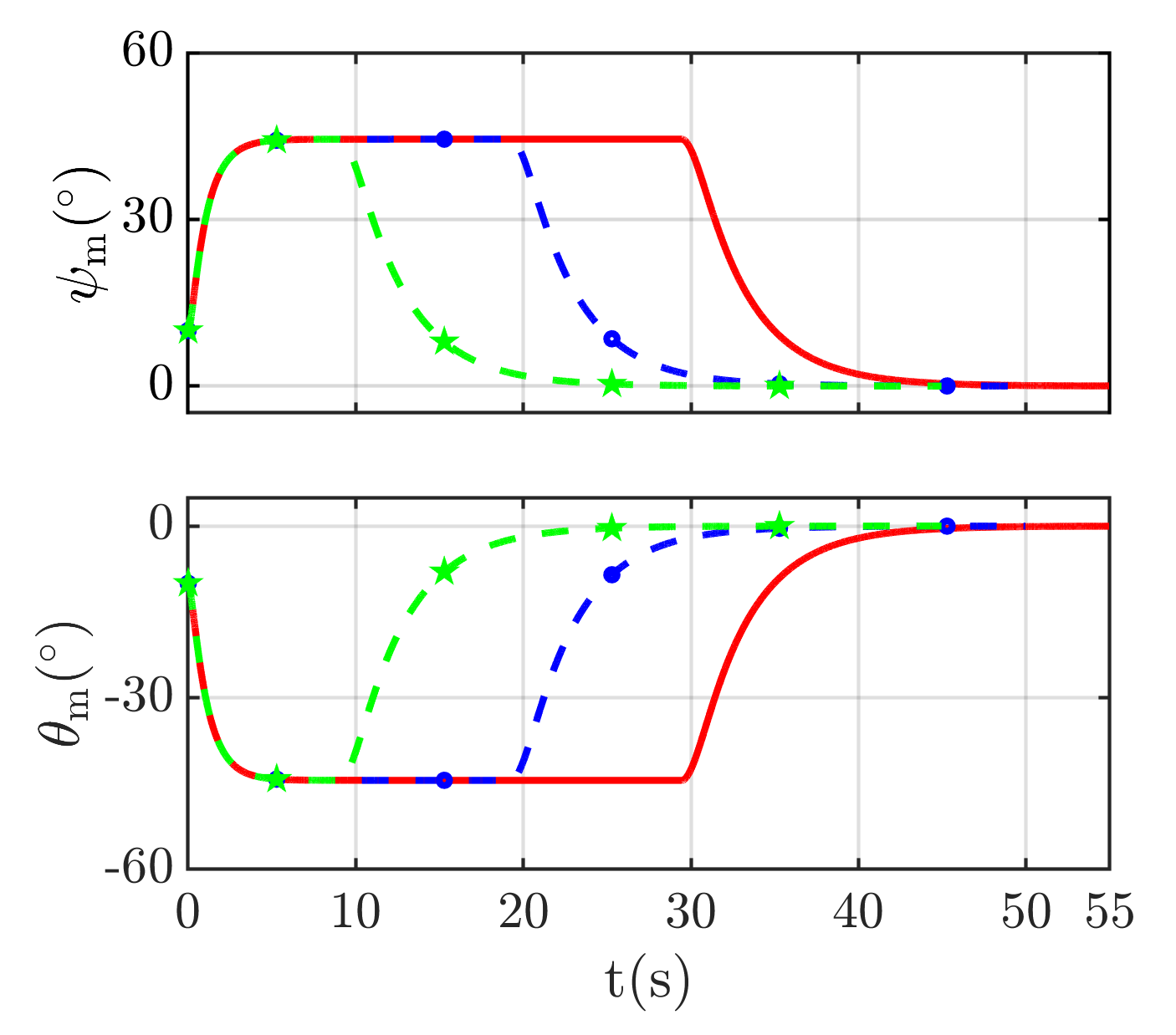}
		\caption{Heading angles.}\label{fig:T10_diff_tf_psi_m_theta_m}
	\end{subfigure}
	\begin{subfigure}[b]{0.245\linewidth}
		\includegraphics[width=\linewidth]{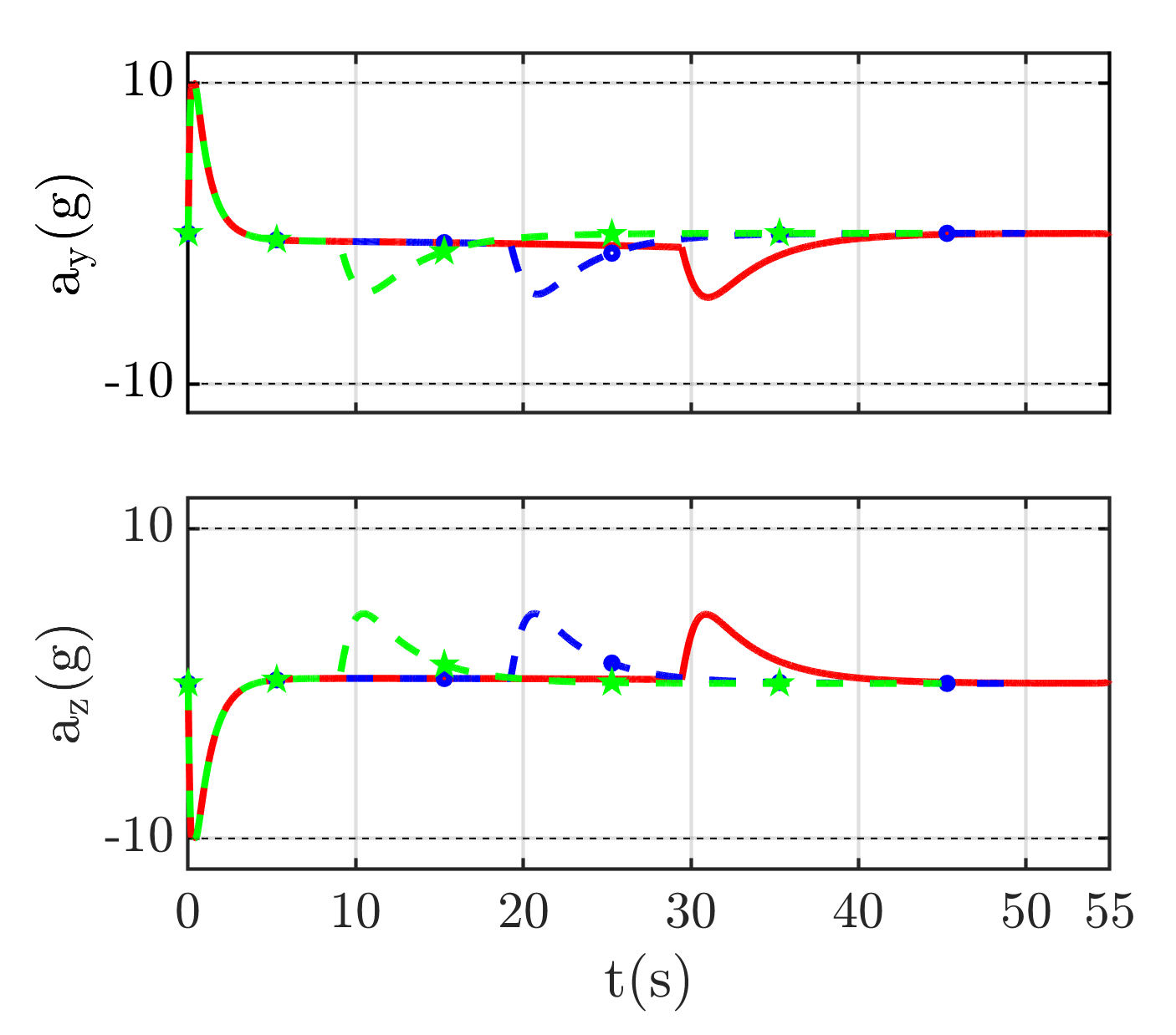}
		\caption{Lateral accelerations.}\label{fig:T10_diff_tf_ay_az}
	\end{subfigure}
	\begin{subfigure}[b]{0.245\linewidth}
		\includegraphics[width=\linewidth]{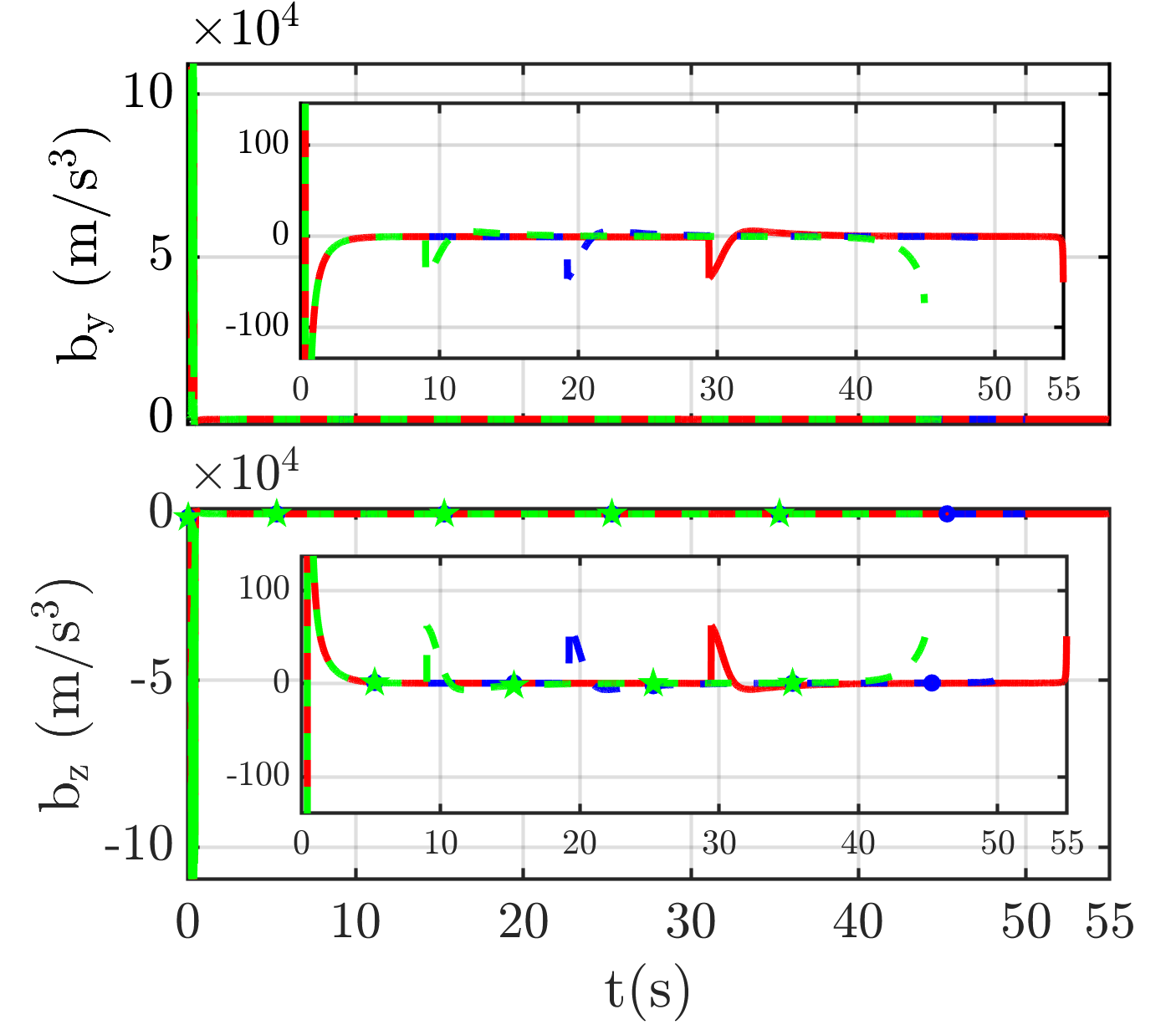}
		\caption{Commanded acceleration.}\label{fig:T10_diff_tf_bybz}
	\end{subfigure}
	\begin{subfigure}[b]{0.245\linewidth}
		\includegraphics[width=\linewidth]{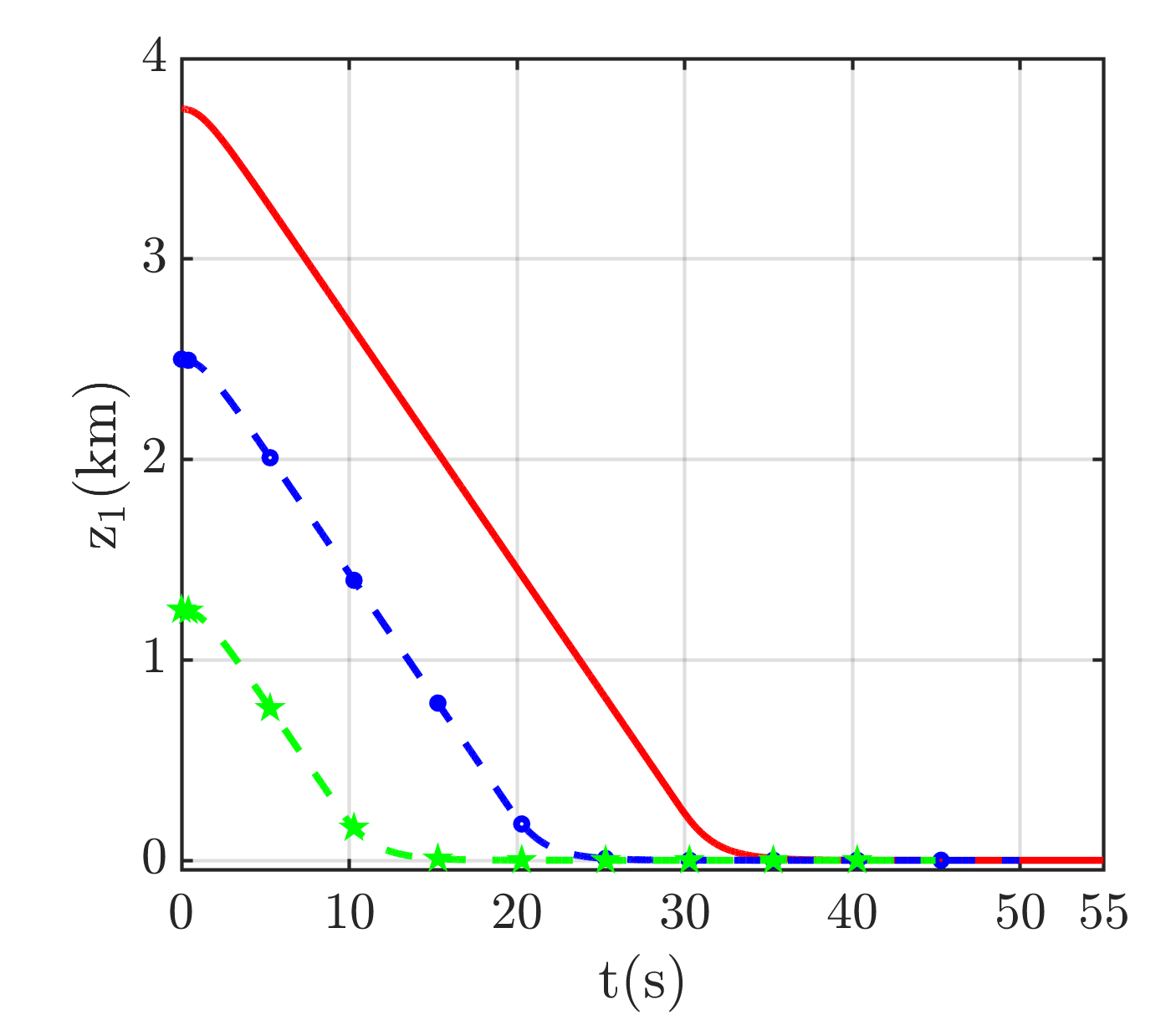}
		\caption{Range error.}\label{fig:T10_diff_tf_z1z2}
	\end{subfigure}
	\begin{subfigure}[b]{0.245\linewidth}
		\includegraphics[width=\linewidth]{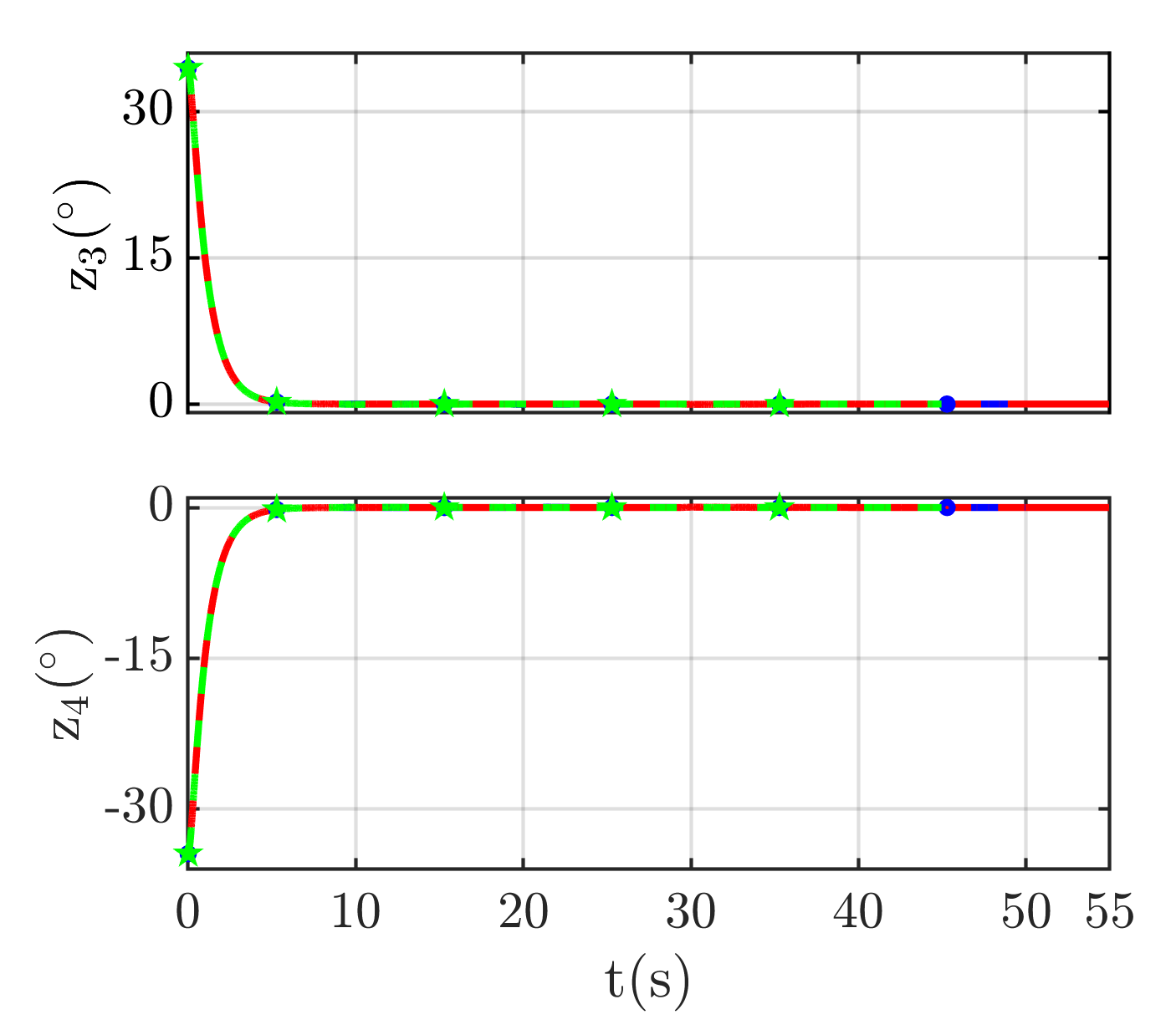}
		\caption{Heading angle errors.}\label{fig:T10_diff_tf_z3_z4}
	\end{subfigure}   
	\begin{subfigure}[b]{0.245\linewidth}
		\includegraphics[width=\linewidth]{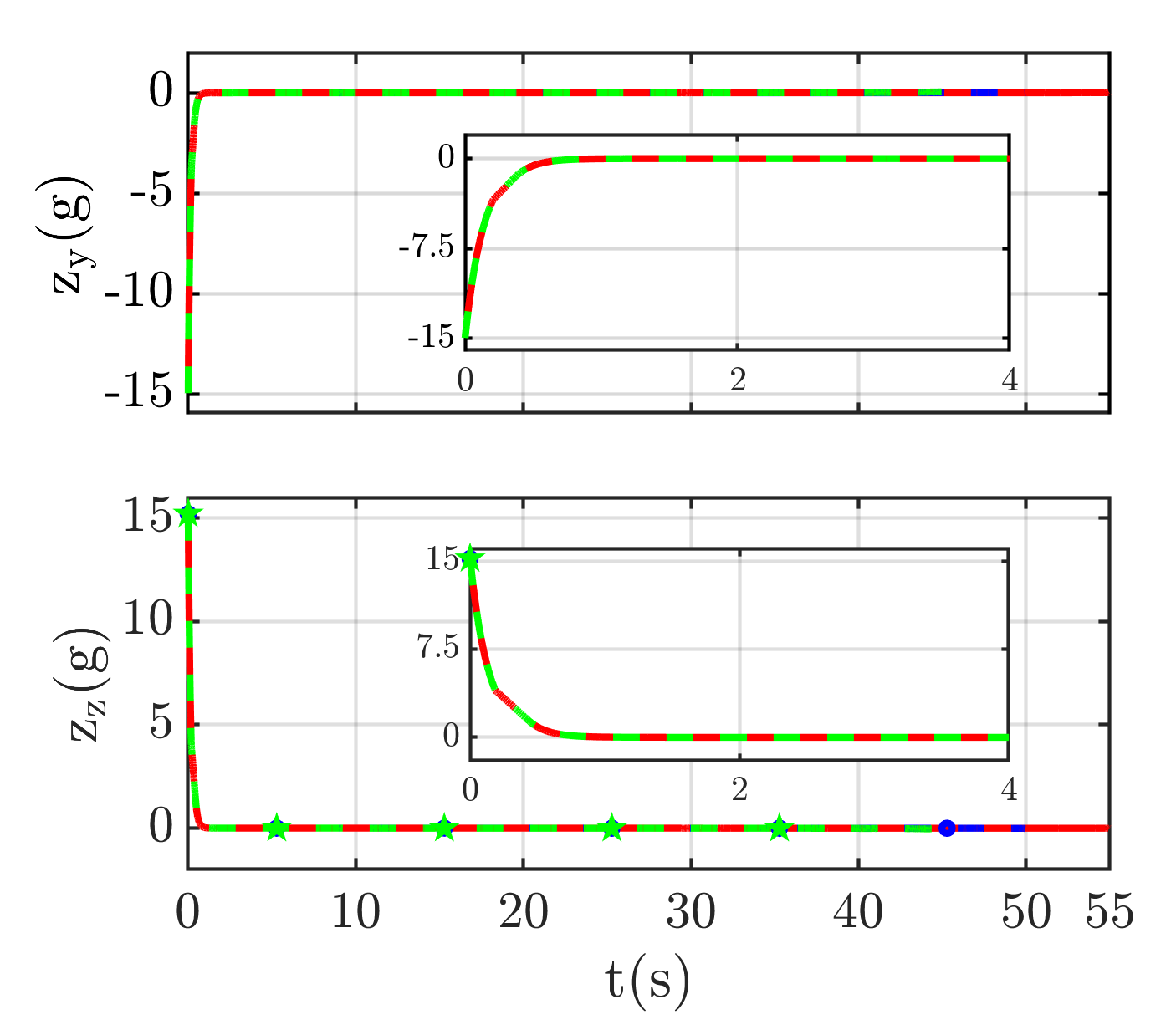}
		\caption{Acceleration errors.}\label{fig:T10_diff_tf_zy_zz}
	\end{subfigure}
	\caption{Performance for different $t_{\rm f}$ and initial heading of $(\theta_\m (0), \psi_\m (0)) = (-10,10)^\circ$.}
	\label{fig:T10_diff_tf}
\end{figure*}

In what follows, we show the effectiveness of the proposed guidance law in \Cref{Sec3B:3D} for 3D engagement scenarios. The simulation results were obtained for impact times of $45$, $50$, and $55\,$s with initial headings, $(\theta_\m(0),\psi_\m(0))$ being $(-10^\circ, 10^\circ)$. The parameters $k_3$, $k_4$, $k_{\rm y}$ and $k_{\rm z}$ are chosen as $1$, $1$, $7$, and $7$, respectively, while the remaining simulation parameters are taken from \Cref{tab:T10_3_Sim_params}. The corresponding plots are shown in \Cref{fig:T10_diff_tf}.
A successful target interception at all the desired impact times is achieved while obeying the FOV and input constraints. These results are shown in \Cref{fig:T10_diff_tf_trajectory,fig:T10_diff_tf_ay_az,fig:T10_diff_tf_r_sigma}. A larger impact time leads to a longer trajectory in the first phase, as seen in \Cref{fig:T10_diff_tf_trajectory}. \cref{fig:T10_diff_tf_z1z2,fig:T10_diff_tf_z3_z4} show that the acceleration errors, $z_y$ and $z_z$, converge to zero before the range and heading angle errors converge. This is necessary for the backstepping control design. The choice of the parameters $k_1$, $k_3$, $k_4$, $k_{\rm y}$ and $k_{\rm z}$ play a crucial role in this aspect.
\begin{figure*}[!h]
	\centering
	\begin{subfigure}[b]{0.245\linewidth}
		\includegraphics[width=\linewidth]{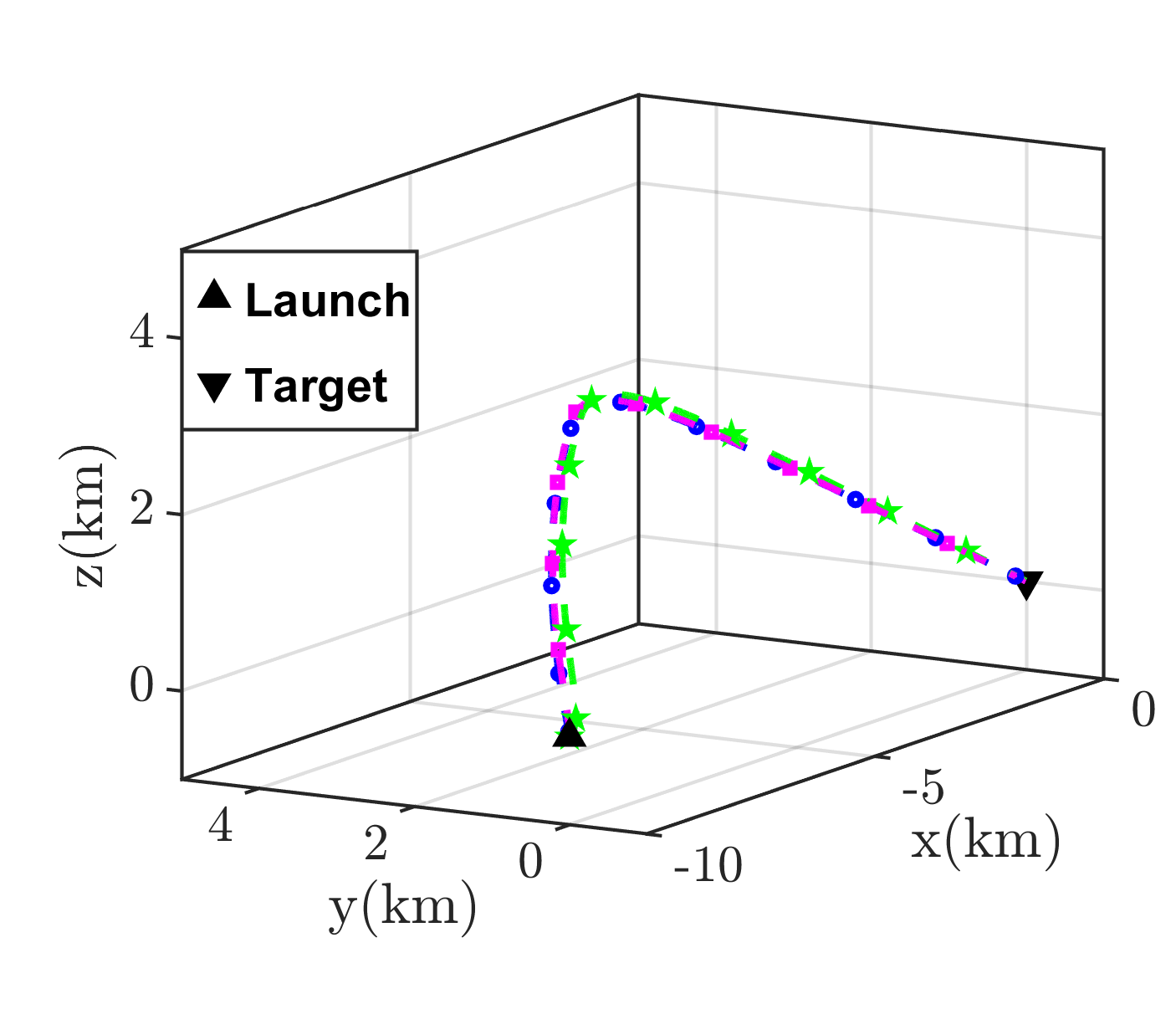}
		\caption{Trajectory.}\label{fig:T10_diff_iniHead_trajectory}
	\end{subfigure}
	\begin{subfigure}[b]{0.245\linewidth}
		\includegraphics[width=\linewidth]{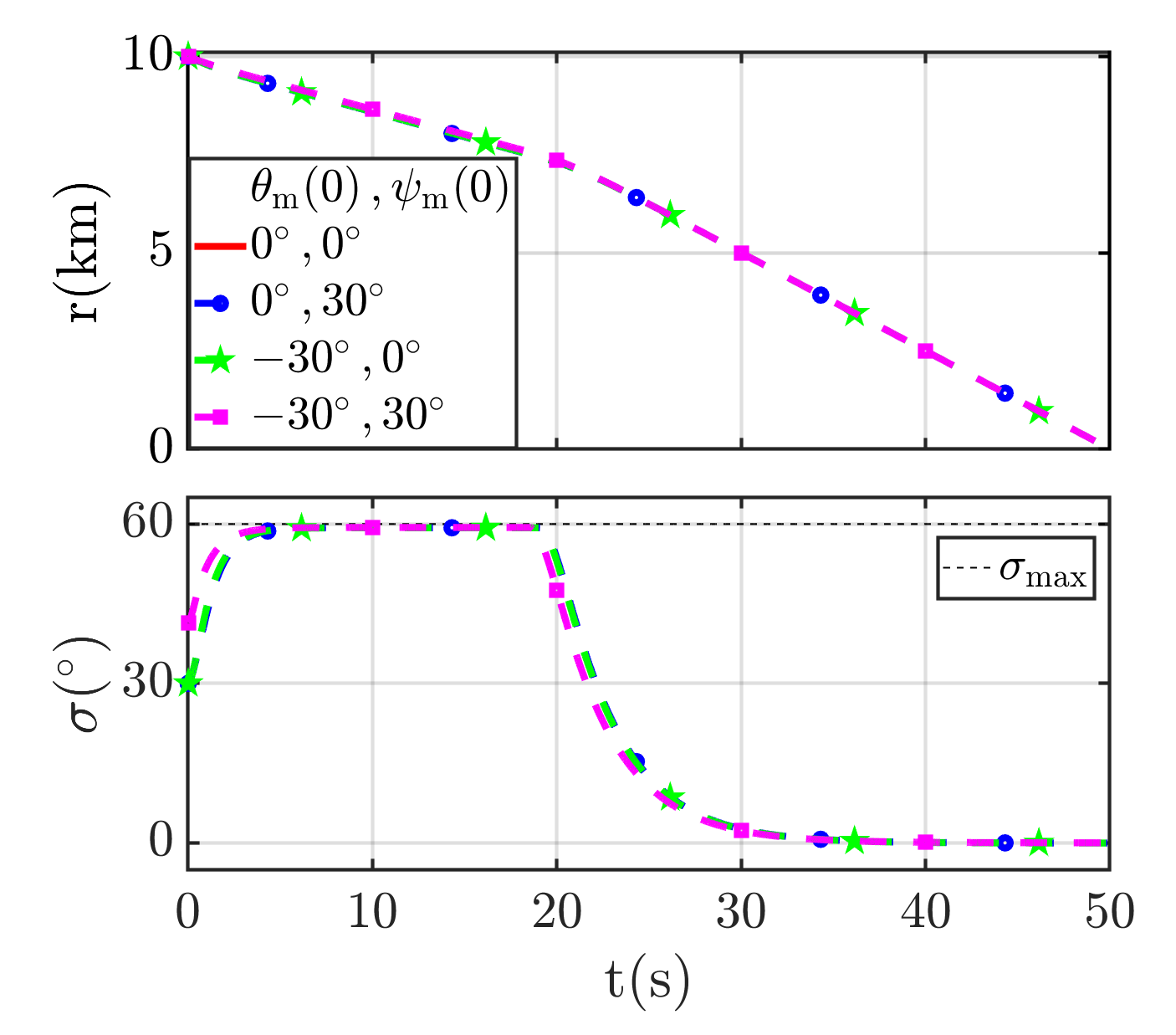}
		\caption{Range and lead angle.}\label{fig:T10_diff_iniHead_r_sigma}
	\end{subfigure}
	\begin{subfigure}[b]{0.245\linewidth}
		\includegraphics[width=\linewidth]{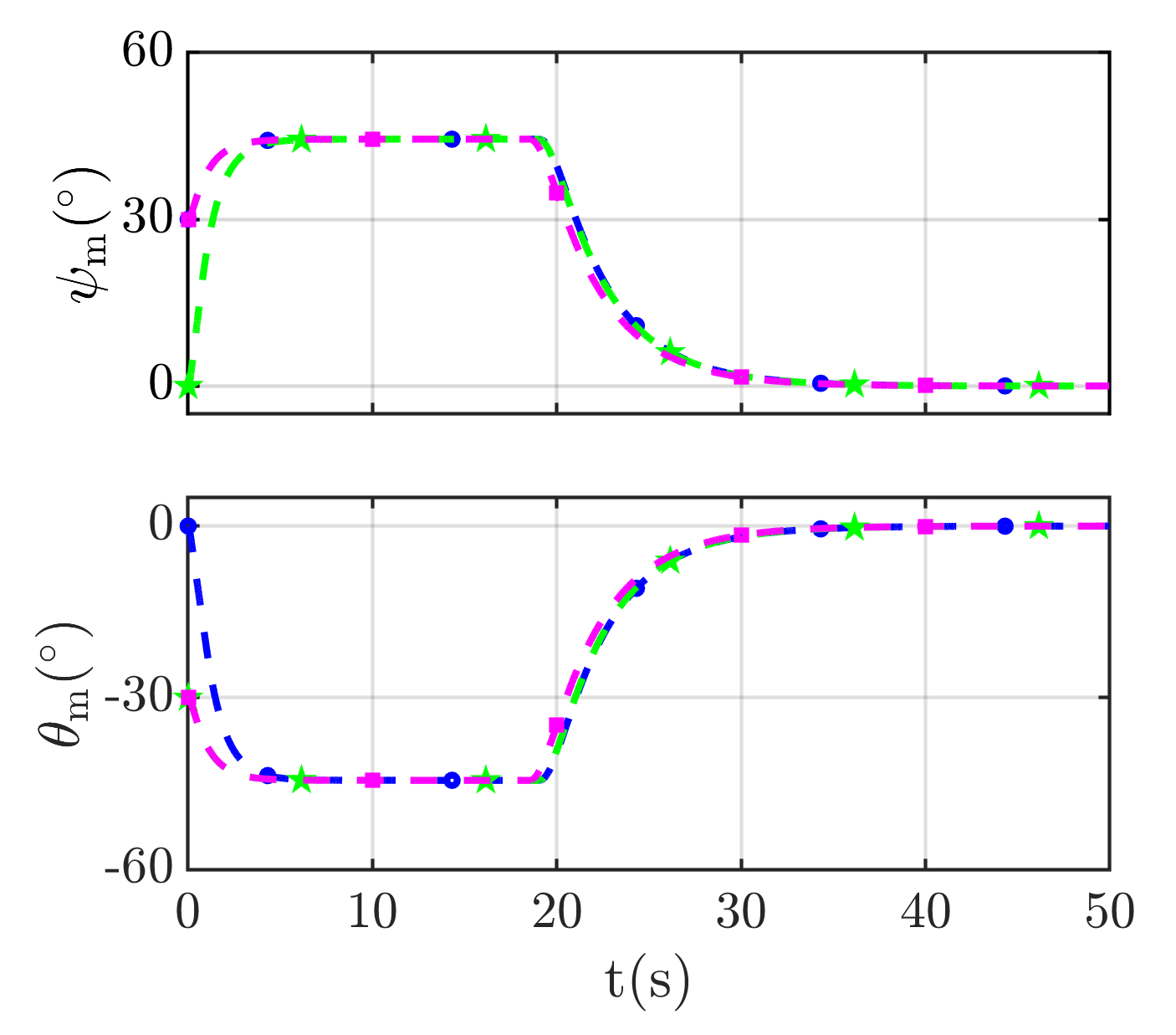}
		\caption{Heading angles.}\label{fig:T10_diff_iniHead_psi_m_theta_m}
	\end{subfigure}
	\begin{subfigure}[b]{0.245\linewidth}
		\includegraphics[width=\linewidth]{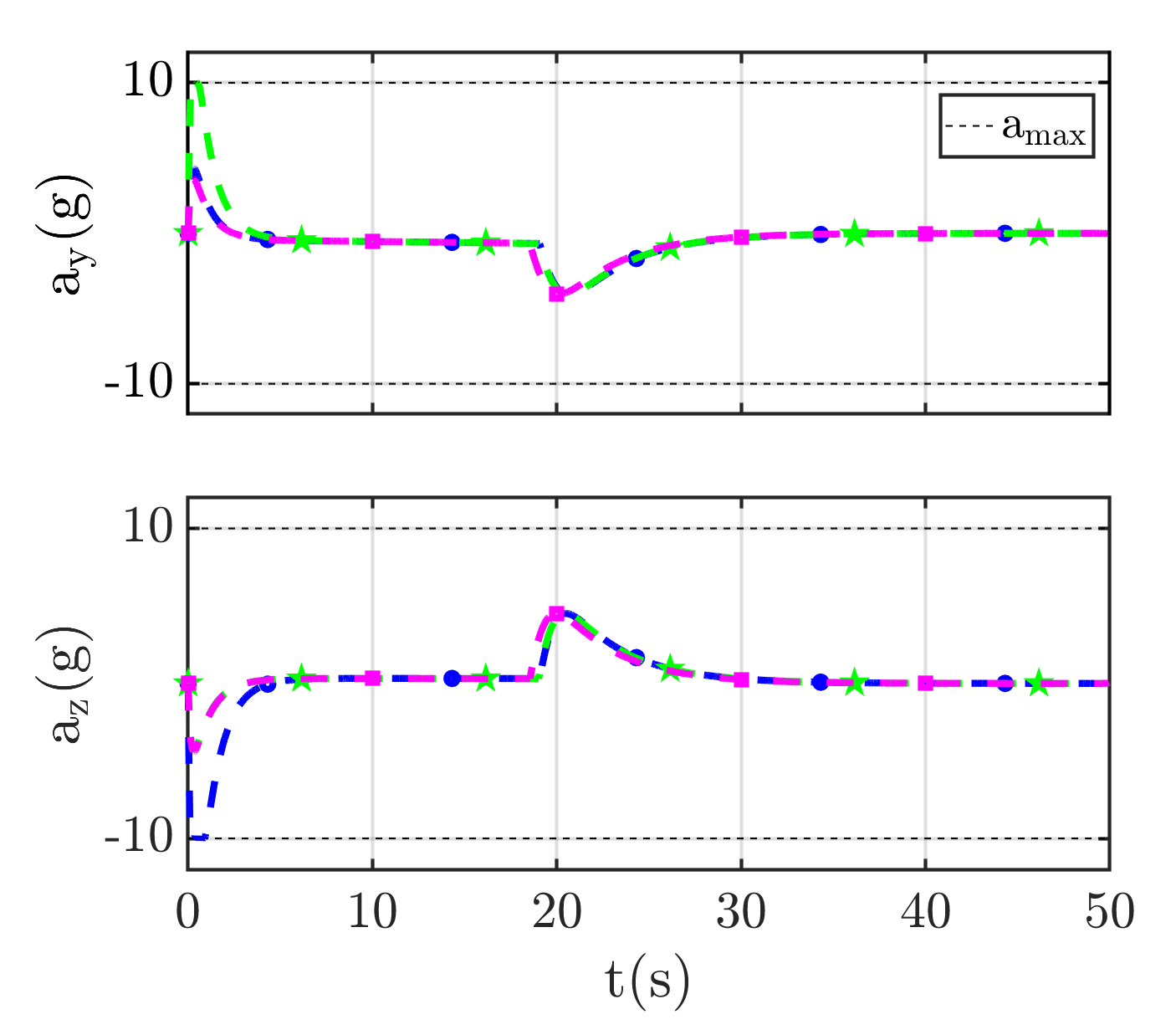}
		\caption{Lateral accelerations.}\label{fig:T10_diff_iniHead_ay_az}
	\end{subfigure}
	\begin{subfigure}[b]{0.245\linewidth}
		\includegraphics[width=\linewidth]{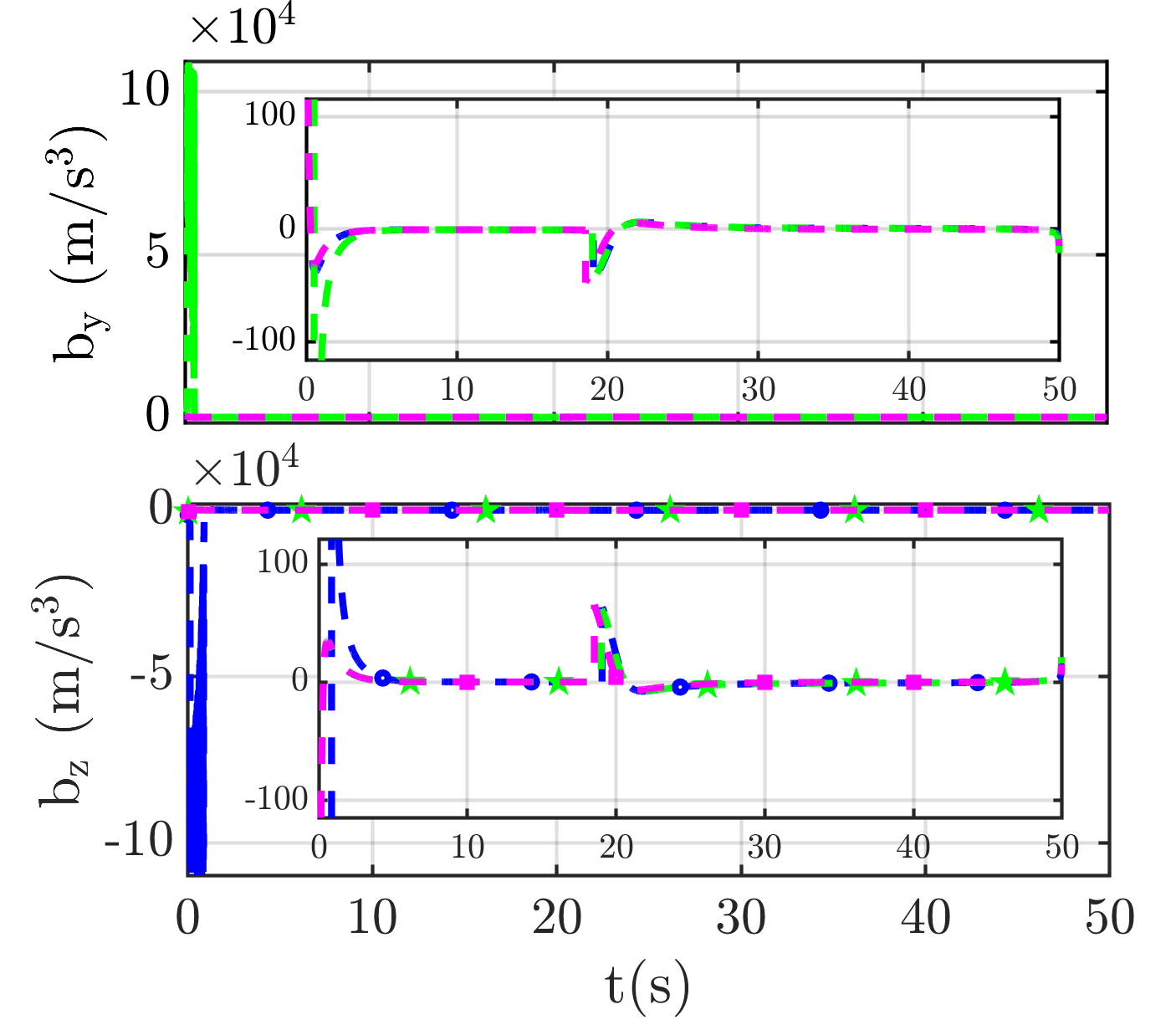}
		\caption{Commanded acceleration.}\label{fig:T10_diff_iniHead_bybz}
	\end{subfigure}
	\begin{subfigure}[b]{0.245\linewidth}
		\includegraphics[width=\linewidth]{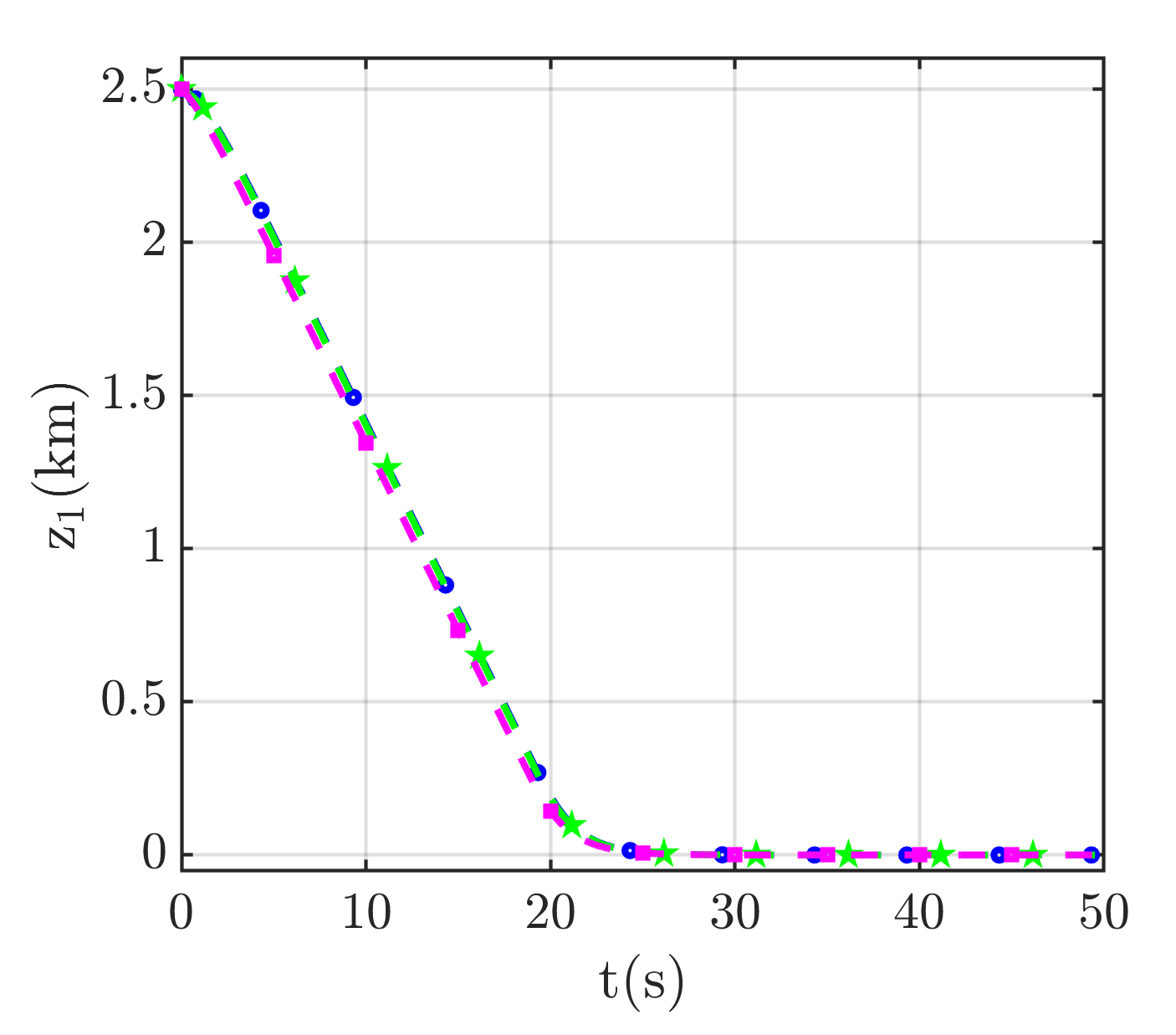}
		\caption{Range error.}\label{fig:T10_diff_iniHead_z1z2}
	\end{subfigure}
	\begin{subfigure}[b]{0.245\linewidth}
		\includegraphics[width=\linewidth]{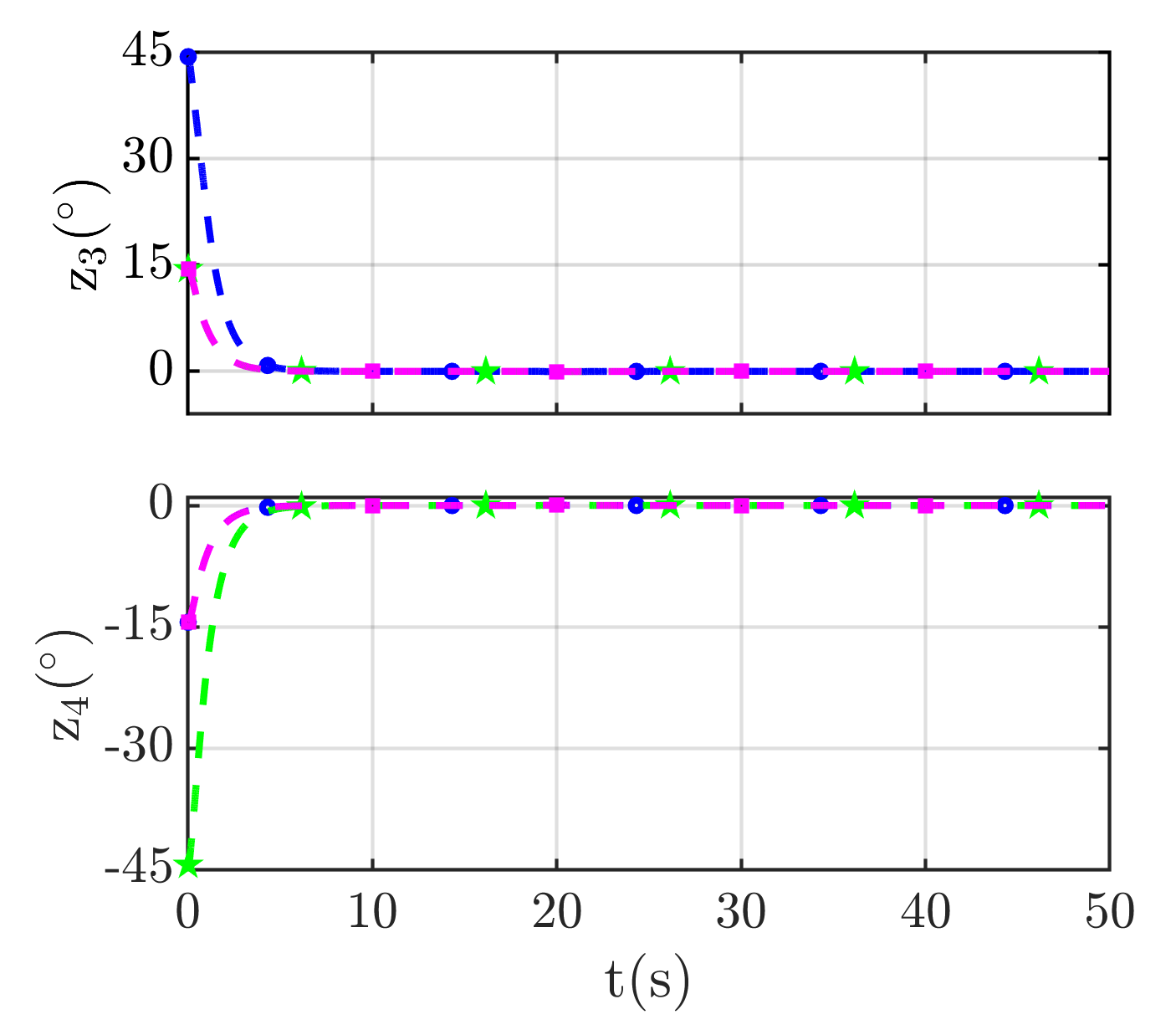}
		\caption{Heading angle errors.}\label{fig:T10_diff_iniHead_z3_z4}
	\end{subfigure}
	\begin{subfigure}[b]{0.245\linewidth}
		\includegraphics[width=\linewidth]{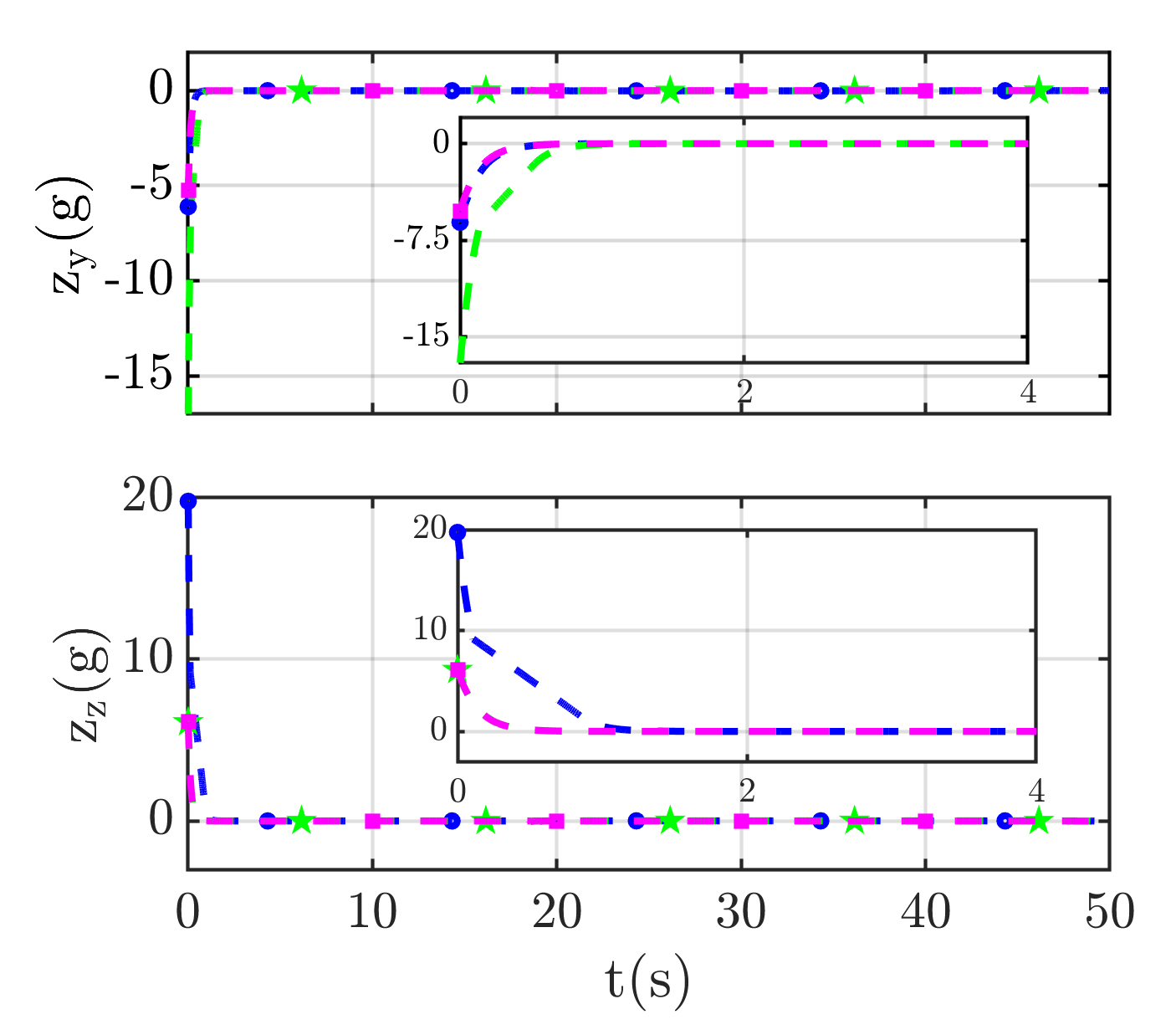}
		\caption{Acceleration errors.}\label{fig:T10_diff_iniHead_zy_zz}
	\end{subfigure}
	\caption{Performance for $t_{\rm f}=50\,$s and different $(\theta_\m (0),\,\psi_\m (0))$.}
	\label{fig:T10_diff_iniHead}
\end{figure*}

Next simulation results of engagement scenarios with initial heading angles $(\theta_\m(0),\,\psi_\m(0))$ values of $(0^\circ,\,0^\circ)$, $(0^\circ,\,30^\circ)$, $(-30^\circ,\,0^\circ)$ and $(-30^\circ,\,30^\circ)$ have been performed the results of which are depicted in  \Cref{fig:T10_diff_iniHead}. From trajectory plots in \Cref{fig:T10_diff_iniHead_trajectory}, one may observe that the trajectory can be split into two phases in correspondence with the virtual inputs chosen in \eqref{eq:T10_bk_step_psi_theta_m_2}. In the first phase, the interceptor maintains non-zero heading angles in order to meet the impact time constraint, while in the second phase, the interceptor moves along the collision course to intercept the target. The proposed guidance law provides additional control over the interceptor's trajectory. The virtual inputs in \eqref{eq:T10_bk_step_psi_theta_m_2} enable the interceptor to maintain the octant of the trajectory. Consequently, as can be seen in \Cref{fig:T10_diff_iniHead_trajectory}, the interceptor takes similar trajectories for different initial heading angles. 
As seen in \Cref{fig:T10_diff_tf_bybz}, 
the commanded input is significantly high. However, the lateral acceleration demand is still well within the specified bound (see \Cref{fig:T10_diff_iniHead_ay_az}). From \eqref{eq:T10_sigma_d_kim}, it follows that with the variation in the parameter $k_1$, the desired lead angle varies. Furthermore, the lateral acceleration is related to the rate of change of the heading angles, which, in turn, is influenced by the values of $k_3$ and $k_4$. Given that the values of $k_1$, $k_3$ and $k_4$ are the same for all the launch angles, lateral acceleration as seen in \Cref{fig:T10_diff_iniHead_ay_az} is high in the transient phase because of the large initial heading angle errors. 
In both scenarios (as shown in \Cref{fig:T10_diff_tf,fig:T10_diff_iniHead}), the acceleration error in \Cref{fig:T10_diff_tf_zy_zz,fig:T10_diff_iniHead_zy_zz} converges to zero well before the range and lead angle errors. In all scenarios, the target is intercepted successfully at the desired impact time while adhering to the FOV as well as input bounds.

\subsection{Performance for 3D engagement with varying acceleration bounds}
\begin{figure*}[!h]
	\centering
	\begin{subfigure}[b]{0.245\linewidth}
		\includegraphics[width=\linewidth]{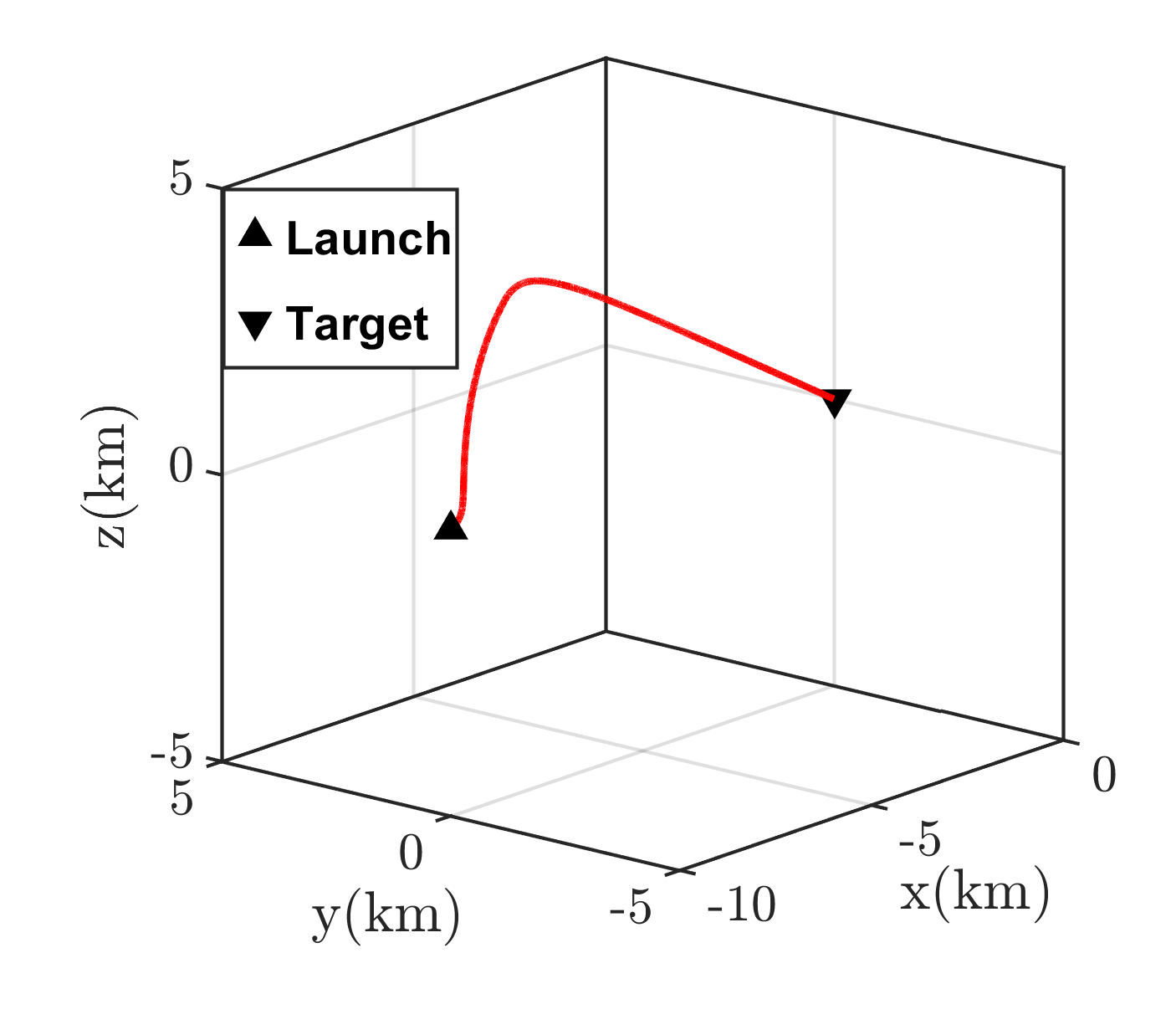}
		\caption{Trajectory.}\label{fig:T10_vary_a_max_l_trajectory}
	\end{subfigure}
	\begin{subfigure}[b]{0.245\linewidth}
		\includegraphics[width=\linewidth]{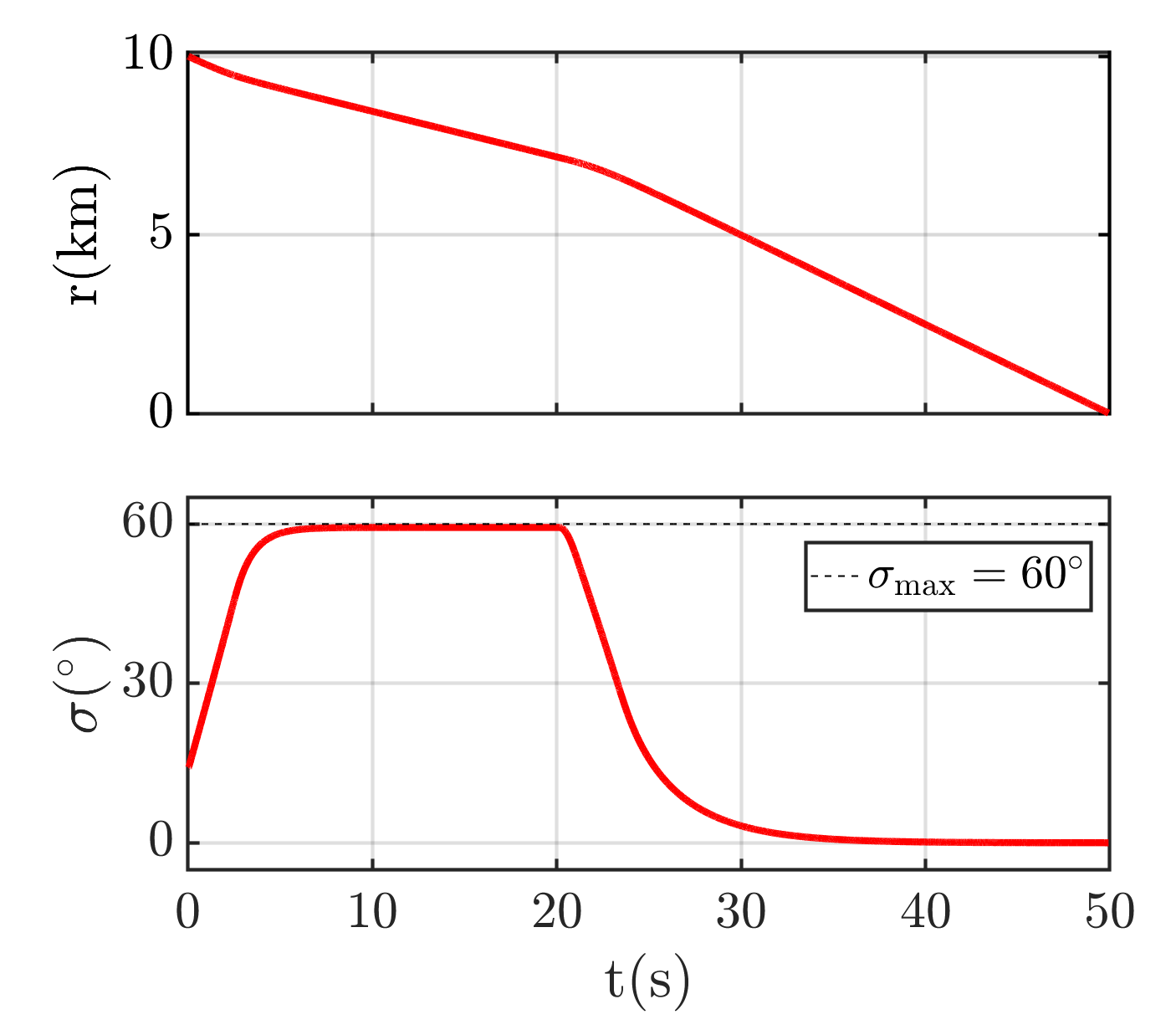}
		\caption{Range and lead angle.}\label{fig:T10_vary_a_max_l_r_sigma}
	\end{subfigure}
	\begin{subfigure}[b]{0.245\linewidth}
		\includegraphics[width=\linewidth]{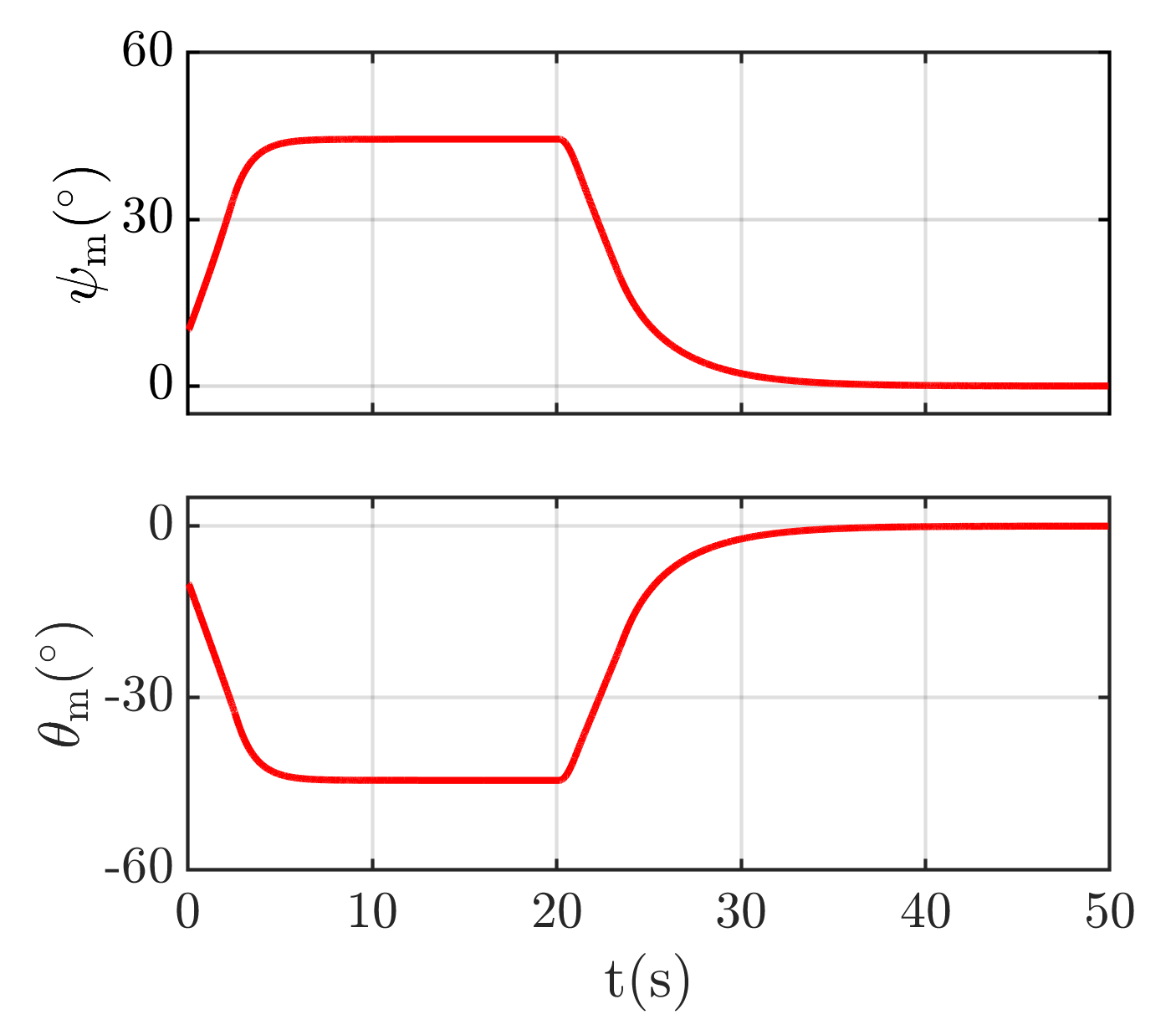}
		\caption{Heading angles.}\label{fig:T10_vary_a_max_l_psi_m_theta_m}
	\end{subfigure}
	\begin{subfigure}[b]{0.245\linewidth}
		\includegraphics[width=\linewidth]{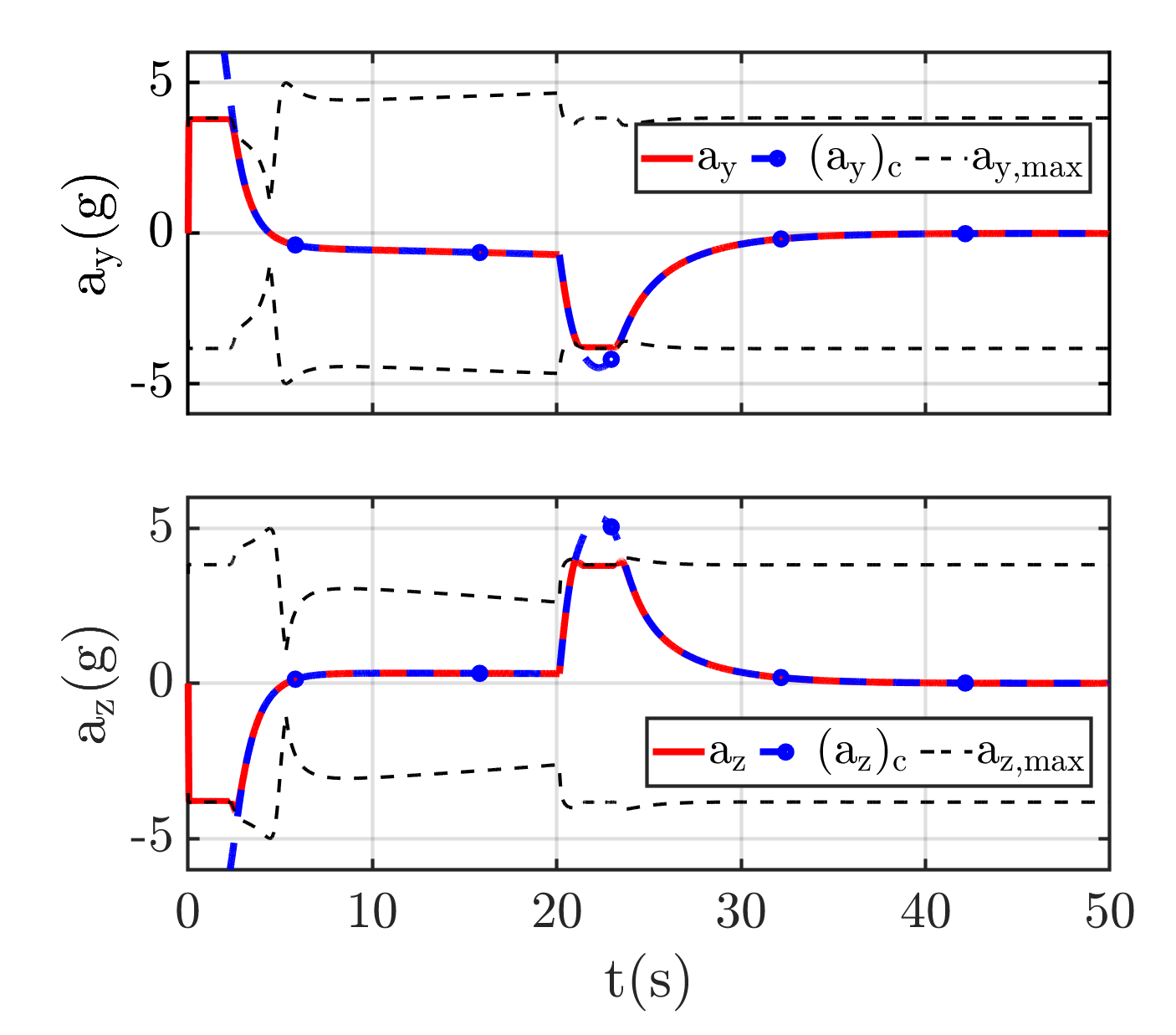}
		\caption{Lateral accelerations.}\label{fig:T10_vary_a_max_l_ay_az}
	\end{subfigure}
	\begin{subfigure}[b]{0.245\linewidth}
		\includegraphics[width=\linewidth]{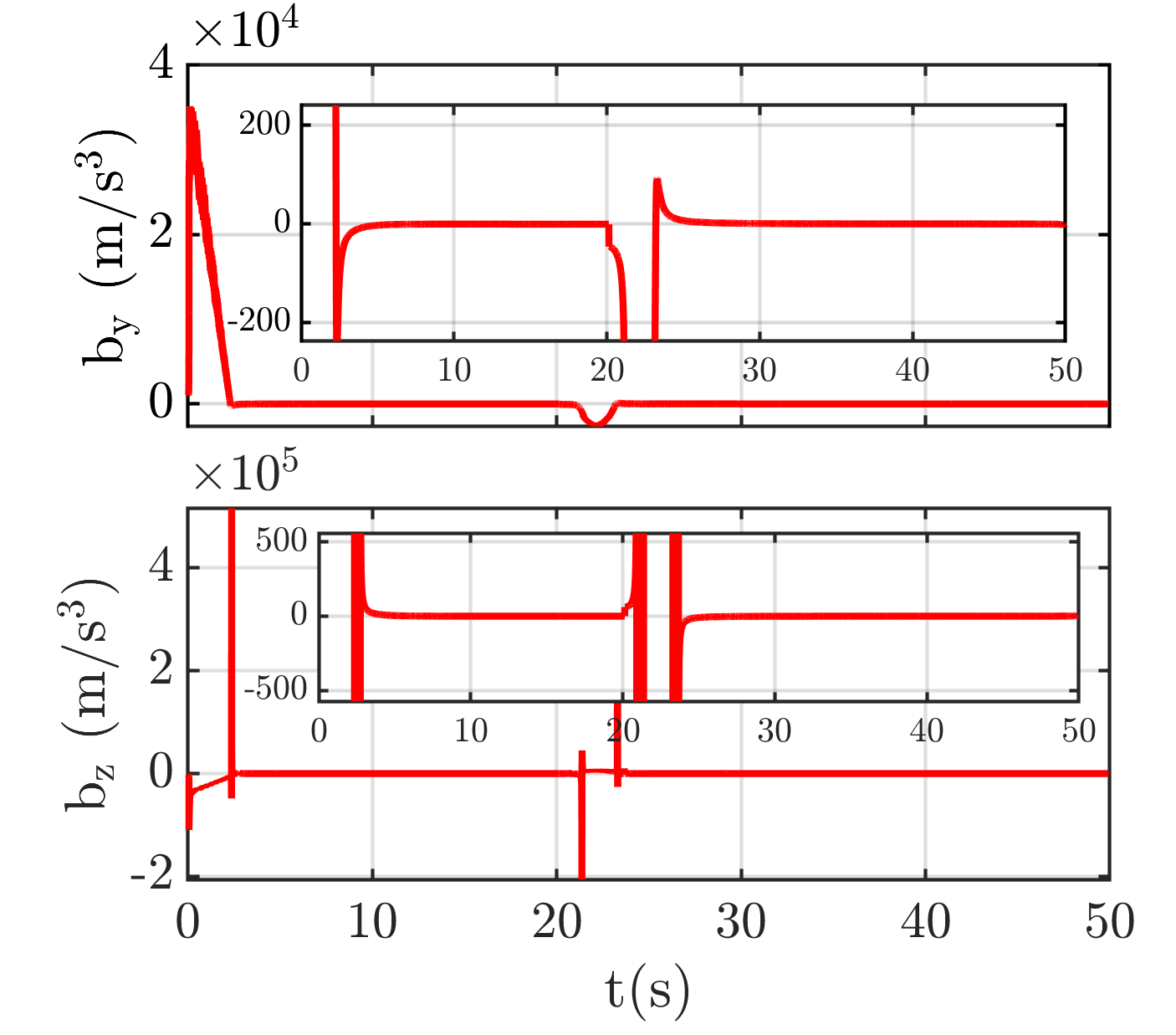}
		\caption{Auxiliary acceleration.}\label{fig:T10_vary_a_max_l_bybz}
	\end{subfigure}
	\begin{subfigure}[b]{0.245\linewidth}
		\includegraphics[width=\linewidth]{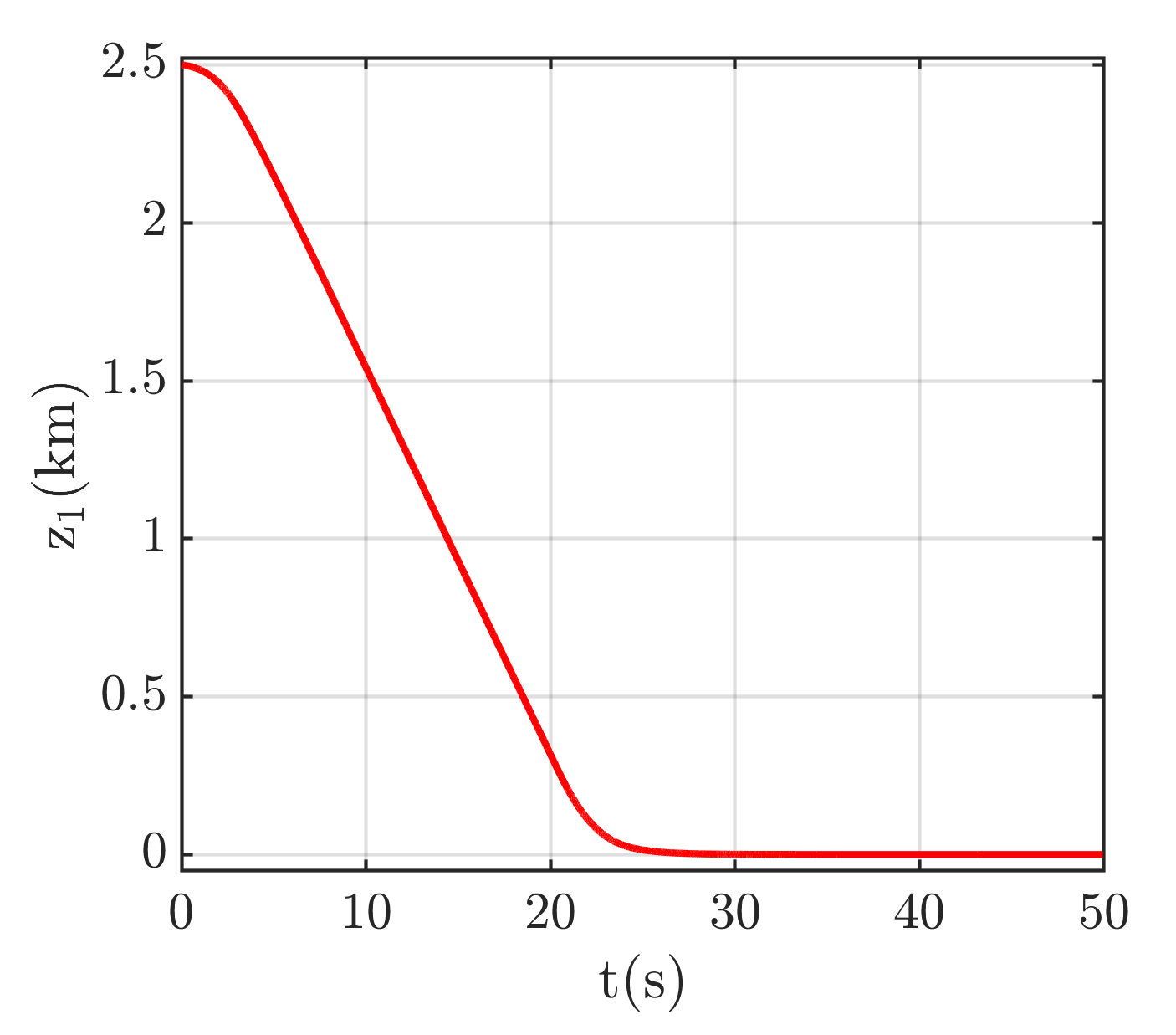}
		\caption{Range error.}\label{fig:T10_vary_a_max_l_z1z2}
	\end{subfigure}
	\begin{subfigure}[b]{0.245\linewidth}
		\includegraphics[width=\linewidth]{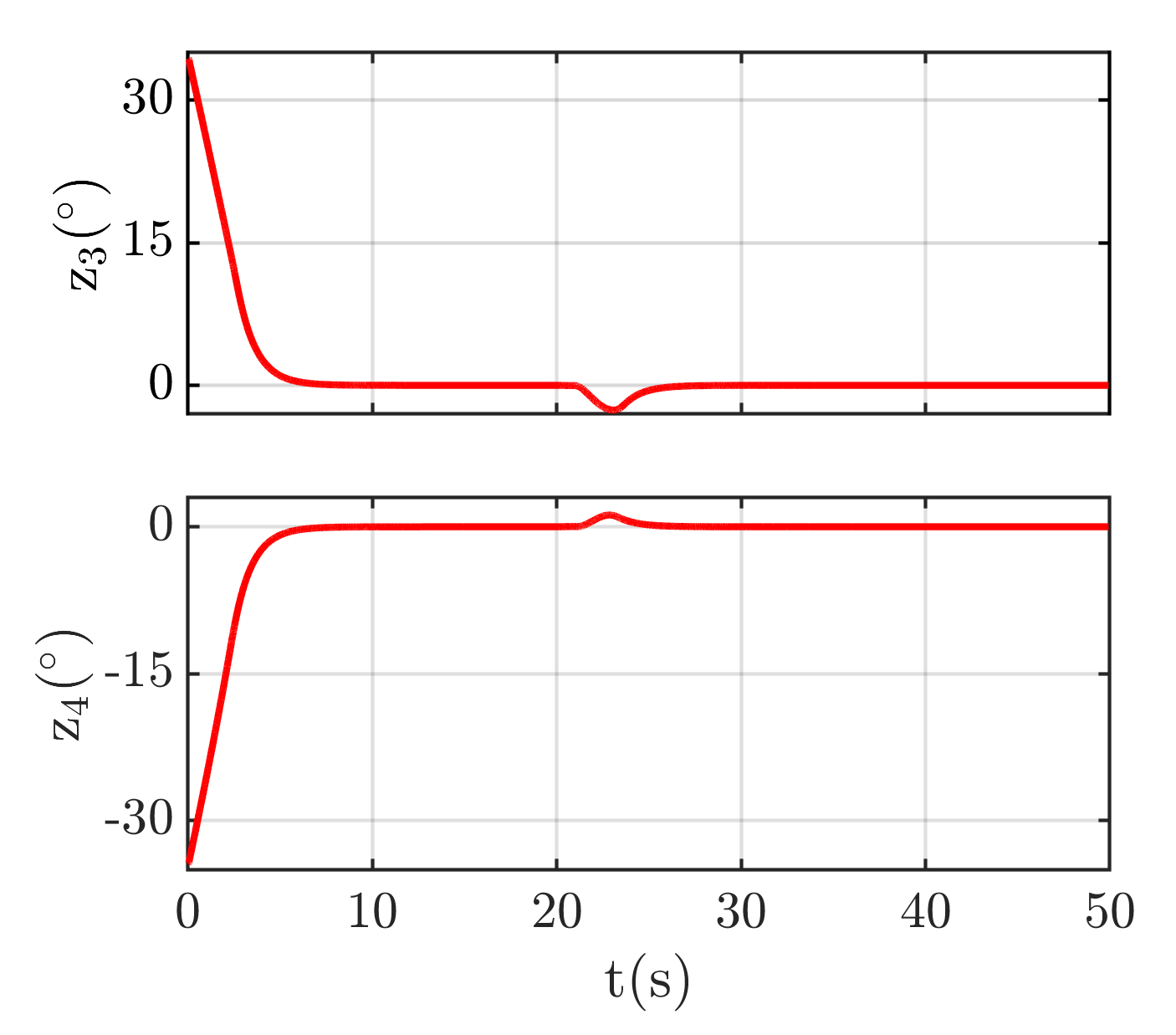}
		\caption{Heading angle errors.}\label{fig:T10_vary_a_max_l_z3_z4}
	\end{subfigure}
	\begin{subfigure}[b]{0.245\linewidth}
		\includegraphics[width=\linewidth]{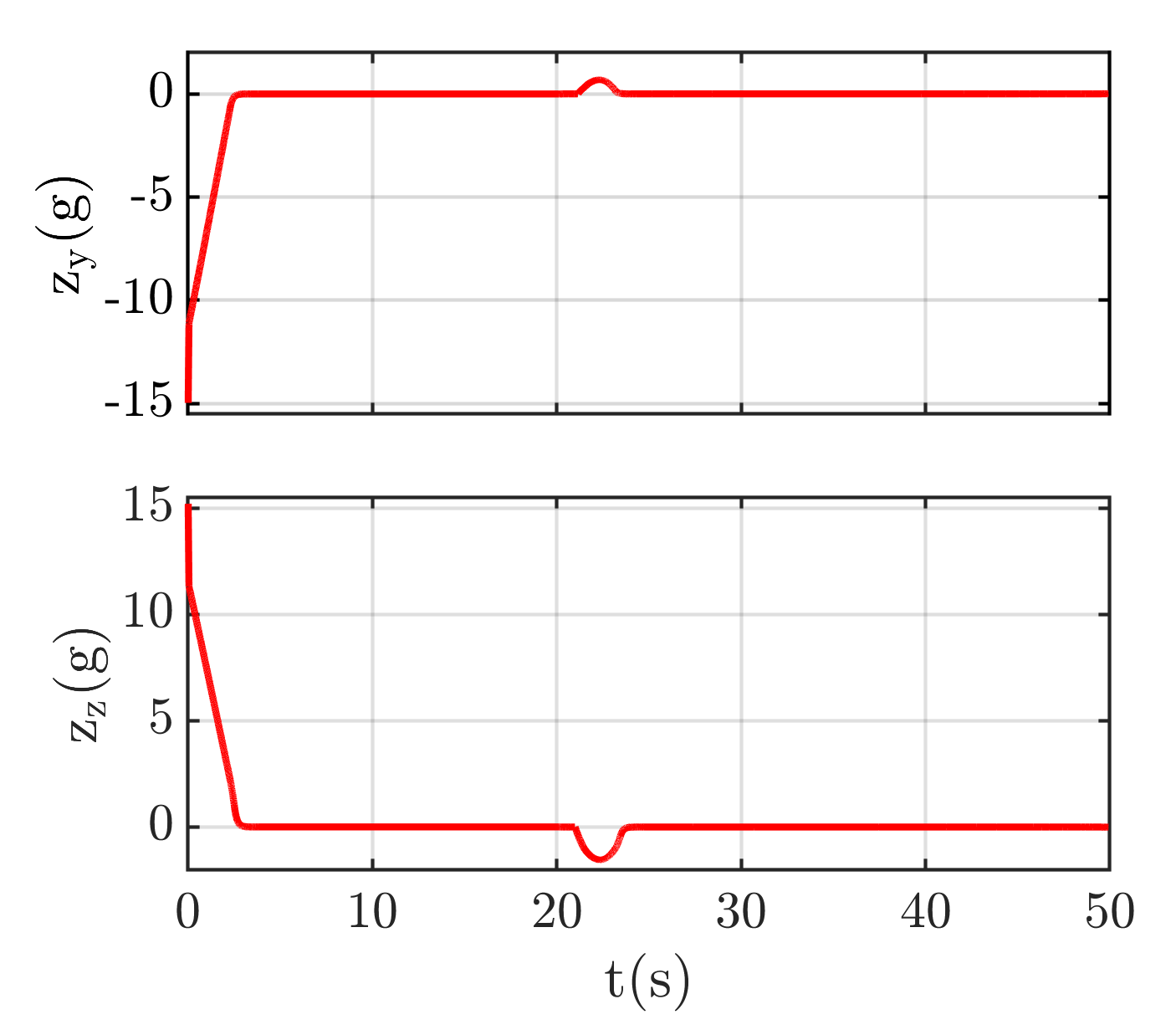}
		\caption{Acceleration errors.}\label{fig:T10_vary_a_max_l_zy_zz}
	\end{subfigure}
	\caption{Performance study with the time-varying acceleration bounds in \eqref{eq:T10_ayMax_azMax_aMax_2}.}
	\label{fig:T10_vary_a_max_l}
\end{figure*}

For the simulation results discussed in the previous sections, the bounds $a_{\rm y,\max}$ and $a_{\rm z,\max}$, were assumed to be constant and equal. However, this assumption is valid only when the interceptors have a plus-type control surface (wing/tail/both) configuration. In what follows next, we have performed simulations with varying values of $a_{\rm y,\max}$ and $a_{\rm z,\max}$. Notably, the maximum acceleration bounds, $a_{\rm y,\max}$ and $a_{\rm z,\max}$, in \eqref{eq:T10_bybz_1} do not necessitate to remain constant throughout the engagement and can vary with time. Therefore, it is imperative to test the performance of the guidance law under realistic, time-varying constraints. For an interceptor with a single pair of radially opposite wings, if the wings are the major contributors to acceleration (through lift production), the interceptor is required to roll in order to produce a significant amount of lateral acceleration, $a_{\rm y}$ and $a_{\rm z}$. This implies that the bound is now on the resultant acceleration and not the individual lateral acceleration components. One may obtain the lateral acceleration components from the resultant acceleration, $a_{\rm r}$, as $ a_{\rm y} = \frac{a_y}{\sqrt{a^2_{\rm y} + a^2_{\rm z}}} a_{\rm r},~~
a_{\rm z} = \frac{a_z}{\sqrt{a^2_{\rm y} + a^2_{\rm z}}} a_{\rm r}$.
Accordingly, one may obtain $a_{\rm y,\max}$ and $a_{\rm z,\max}$ from the maximum resultant acceleration, $a_{\max}$, as follows
\begin{equation}\label{eq:T10_ayMax_azMax_aMax_1}
	a_{\rm y,\max} = \dfrac{a_y}{\sqrt{a^2_{\rm y} + a^2_{\rm z}}} a_{\max},~ 
	a_{\rm z,\max} = \dfrac{a_z}{\sqrt{a^2_{\rm y} + a^2_{\rm z}}} a_{\max}.
\end{equation} 
From \eqref{eq:T10_ayMax_azMax_aMax_1}, one may observe that the terms 
$(a_y/\sqrt{a^2_{\rm y} + a^2_{\rm z}})$, and $(a_z/\sqrt{a^2_{\rm y} + a^2_{\rm z}})$ correspond to the sine and cosine ratios of the interceptor's roll angle. Hence, the bound on the lateral accelerations 
can be interpreted as constraints on the interceptor's roll angle. The simulations are performed assuming a maximum resultant acceleration of $10\,$g. The remaining simulation parameters are taken from \Cref{tab:T10_3_Sim_params}.
\begin{figure*}[!h]
	\centering
	\begin{subfigure}[b]{0.245\linewidth}
		\includegraphics[width=\linewidth]{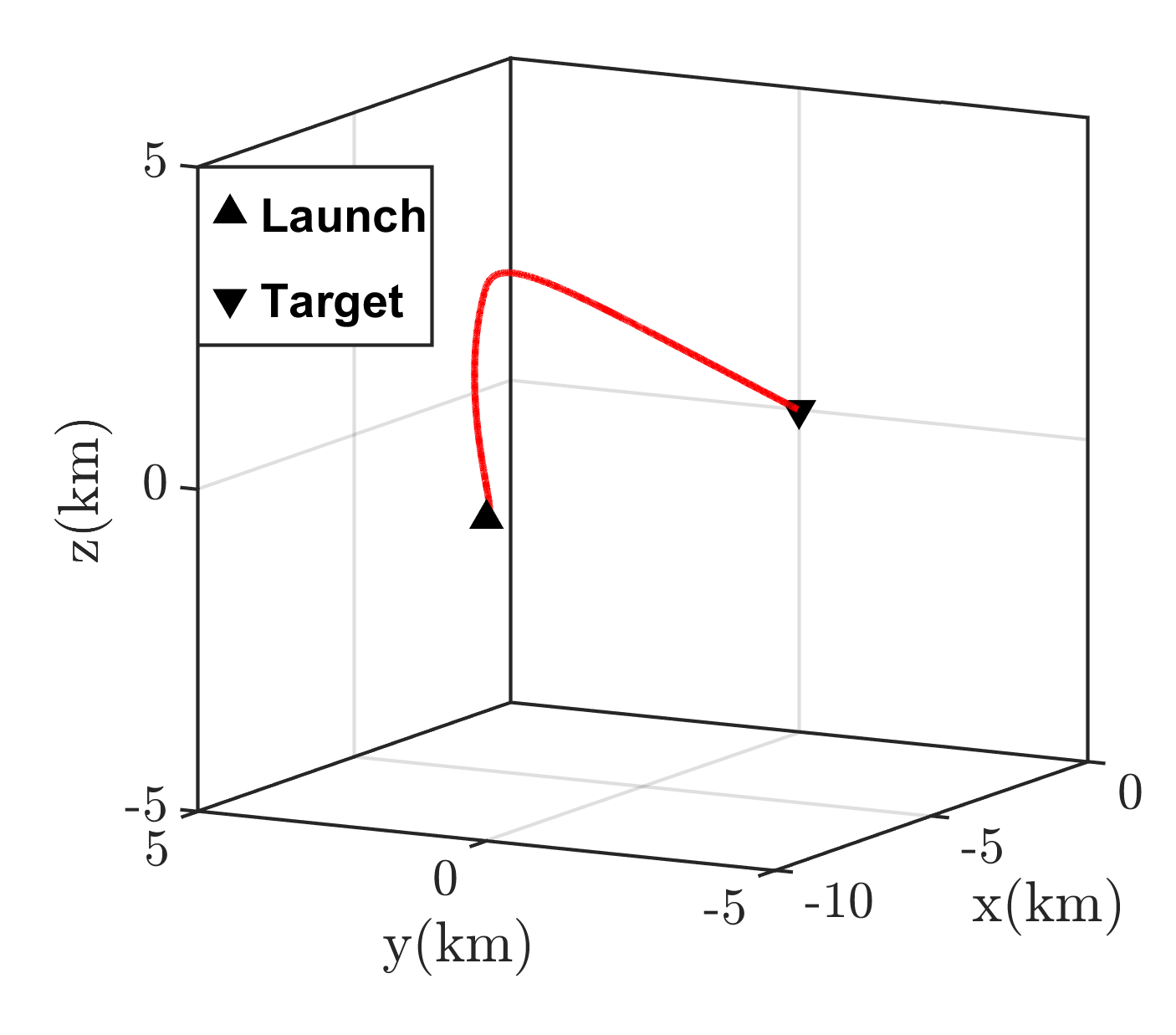}
		\caption{Trajectory.}\label{fig:T10_vary_a_max_trajectory}
	\end{subfigure}
	\begin{subfigure}[b]{0.245\linewidth}
		\includegraphics[width=\linewidth]{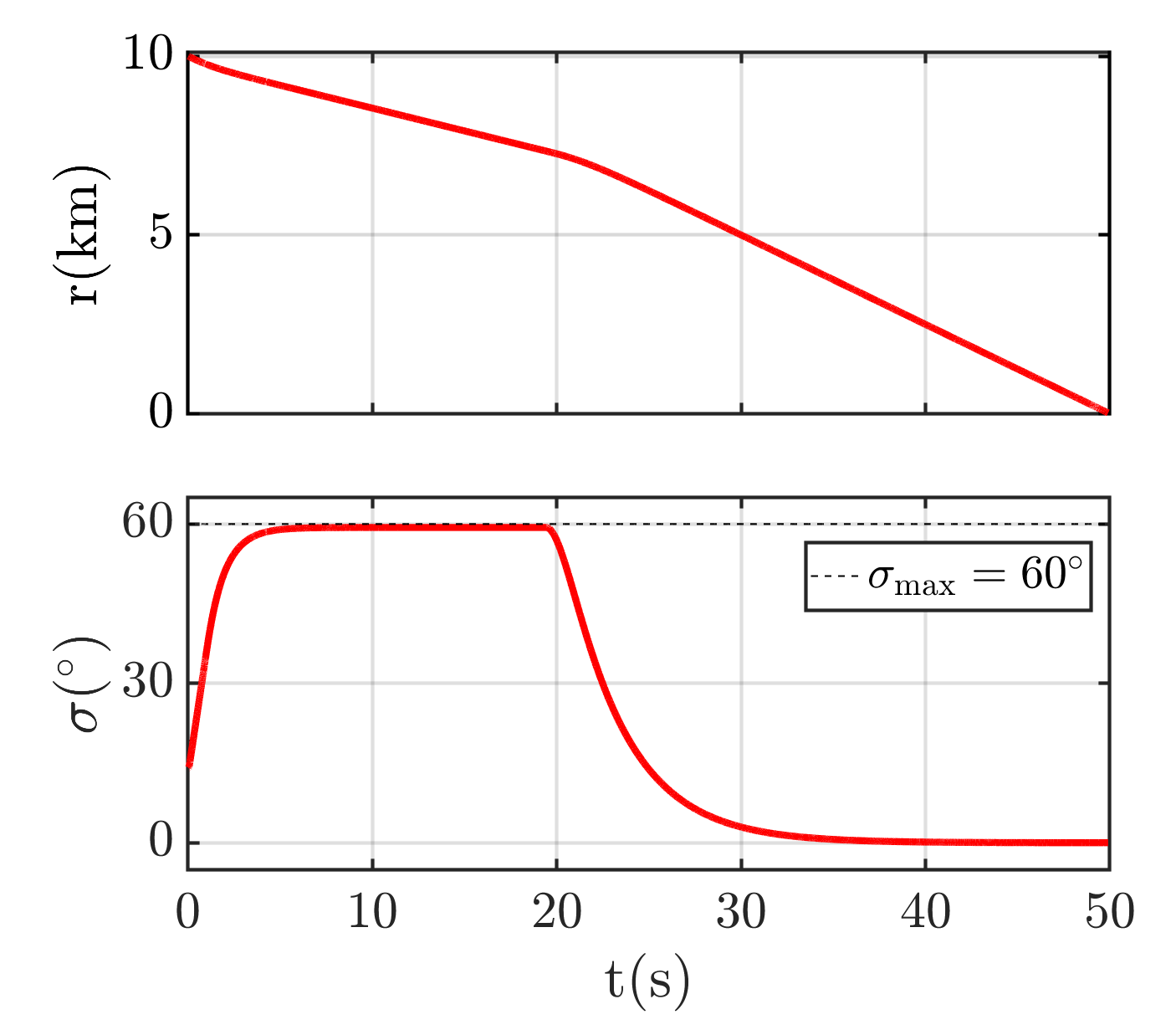}
		\caption{Range and lead angle.}\label{fig:T10_vary_a_max_r_sigma}
	\end{subfigure}
	\begin{subfigure}[b]{0.245\linewidth}
		\includegraphics[width=\linewidth]{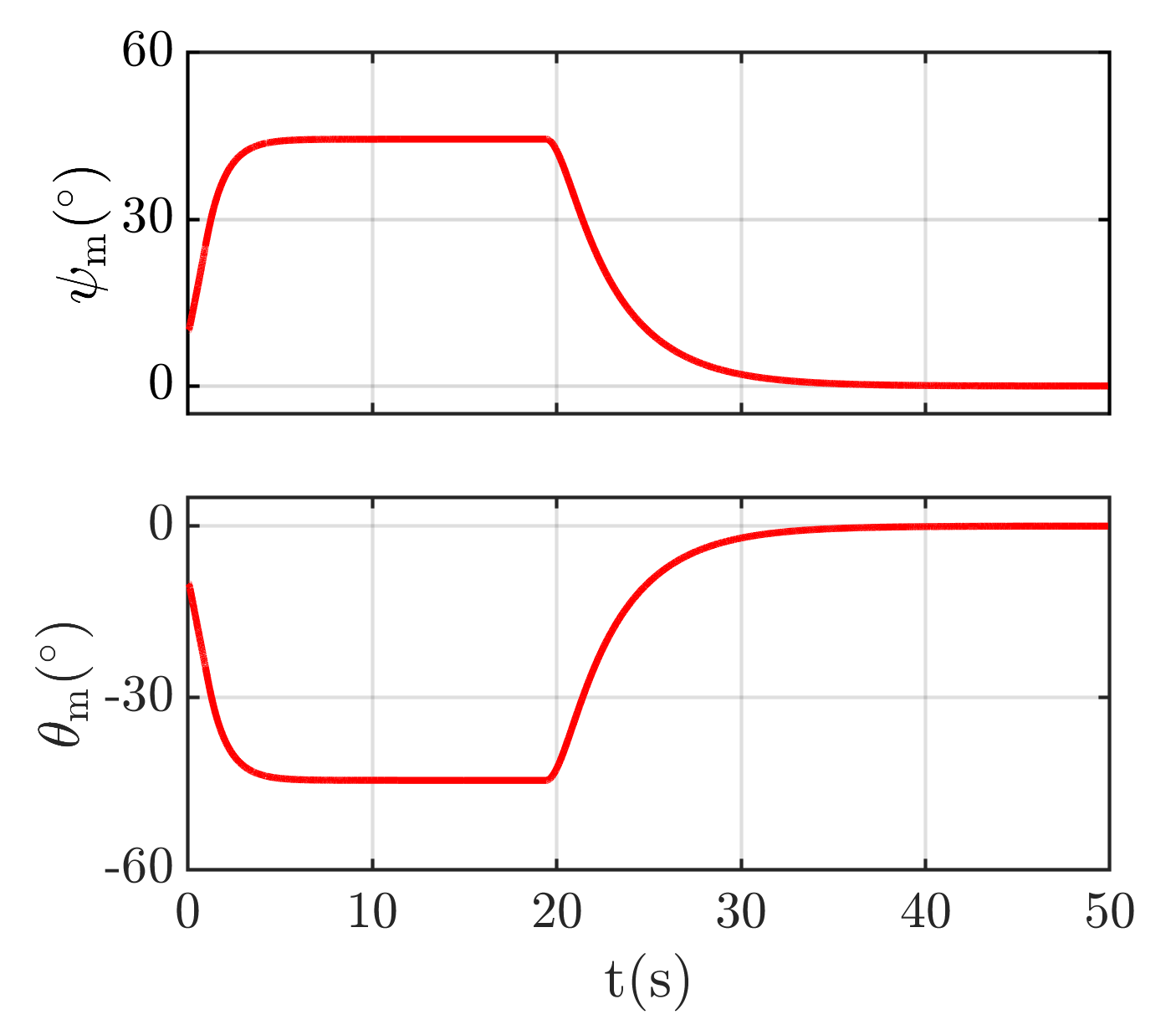}
		\caption{Heading angles.}\label{fig:T10_vary_a_max_psi_m_theta_m}
	\end{subfigure}
	\begin{subfigure}[b]{0.245\linewidth}
		\includegraphics[width=\linewidth]{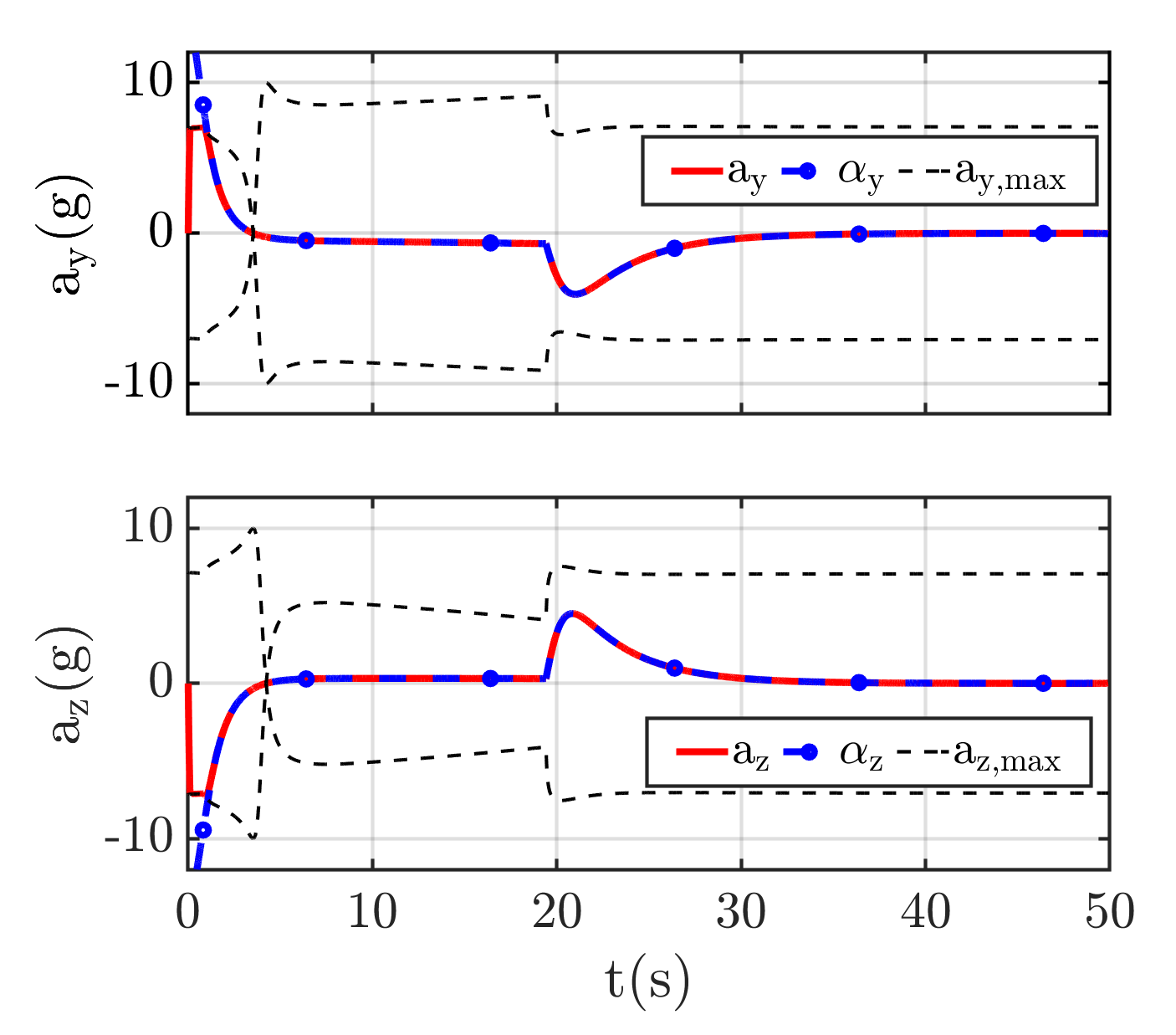}
		\caption{Lateral accelerations.}\label{fig:T10_vary_a_max_ay_az}
	\end{subfigure}
	\begin{subfigure}[b]{0.245\linewidth}
		\includegraphics[width=\linewidth]{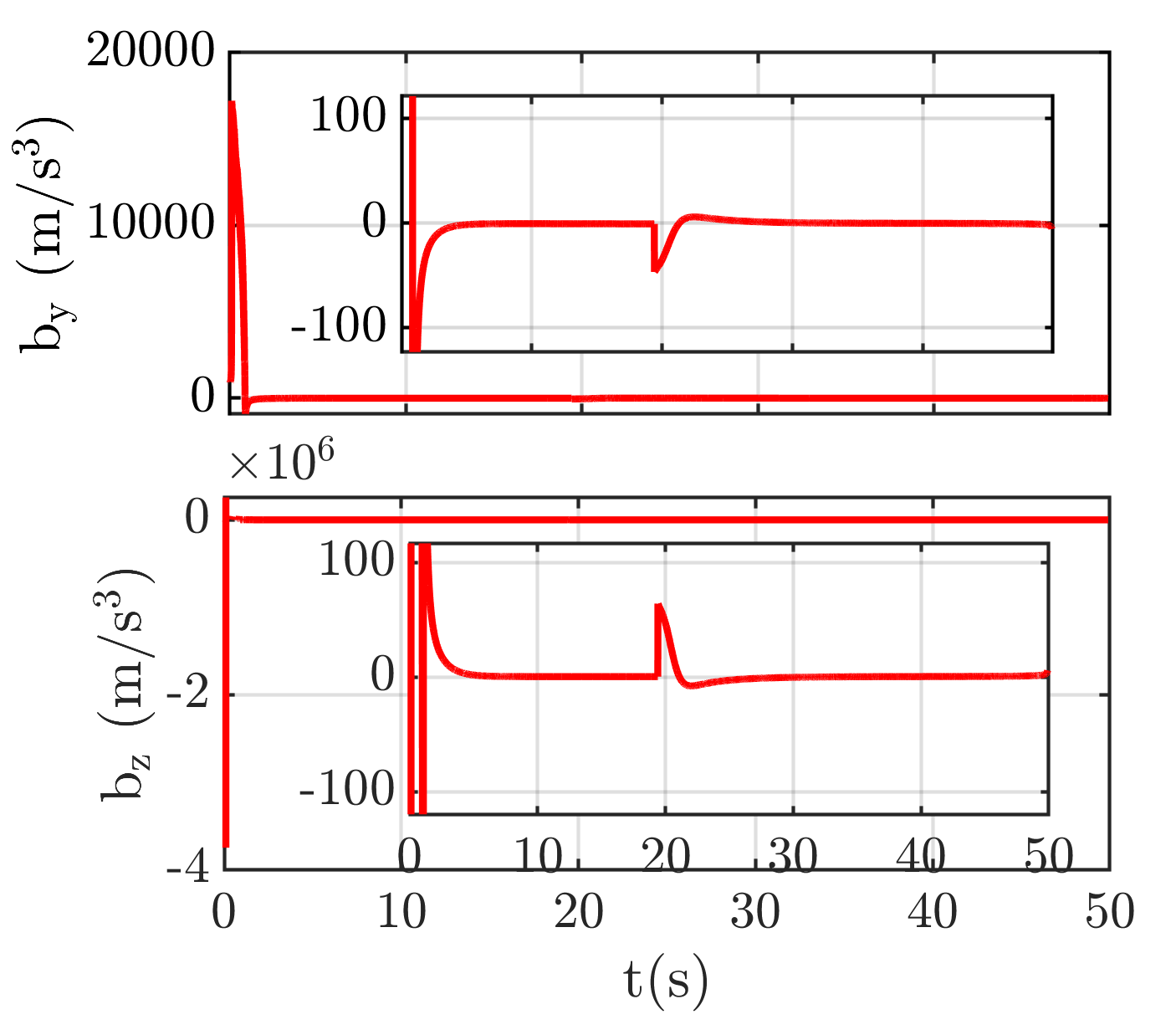}
		\caption{Auxiliary acceleration.}\label{fig:T10_vary_a_max_bybz}
	\end{subfigure}
	\begin{subfigure}[b]{0.245\linewidth}
		\includegraphics[width=\linewidth]{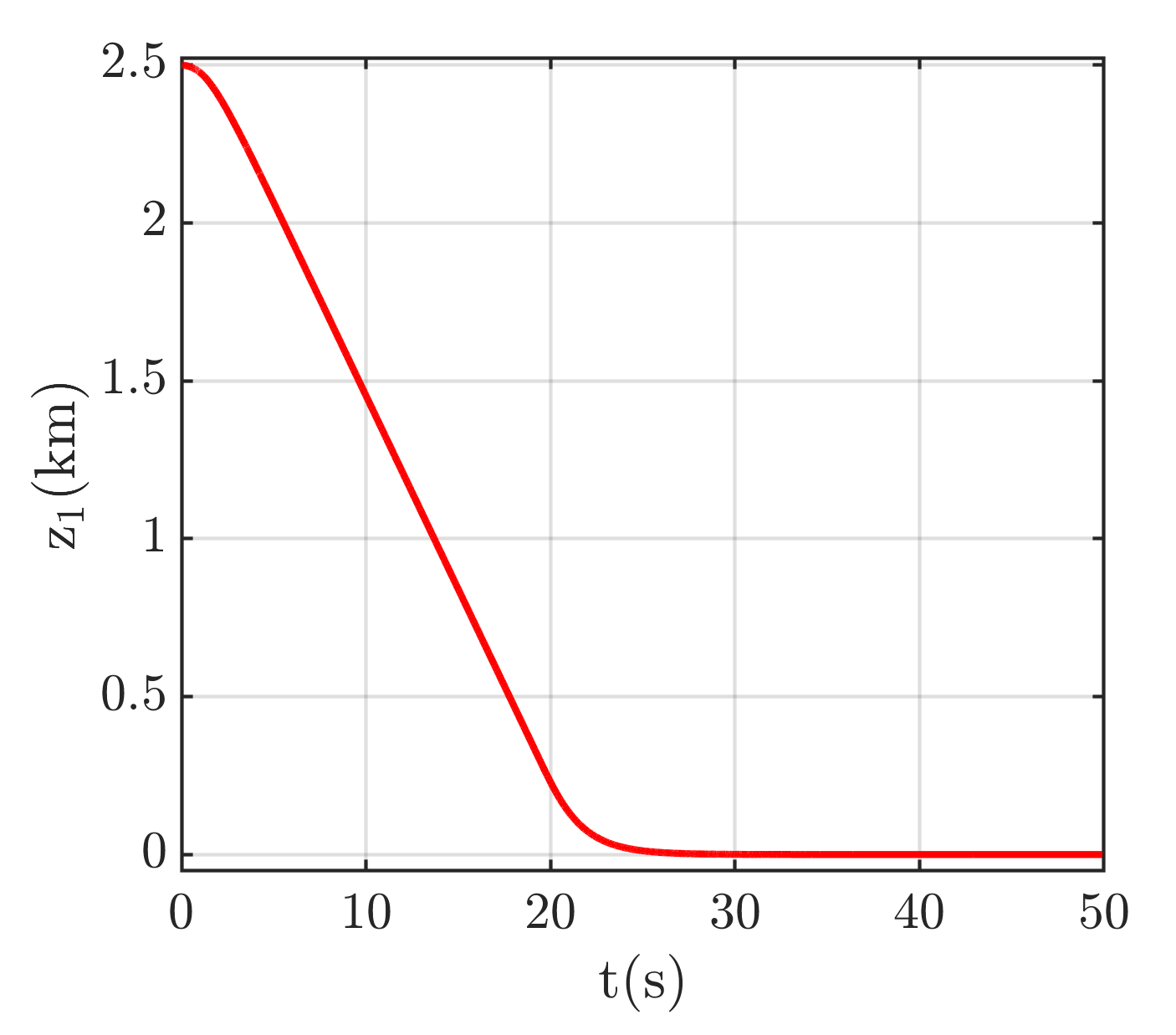}
		\caption{Range error.}\label{fig:T10_vary_a_max_z1z2}
	\end{subfigure}
	\begin{subfigure}[b]{0.245\linewidth}
		\includegraphics[width=\linewidth]{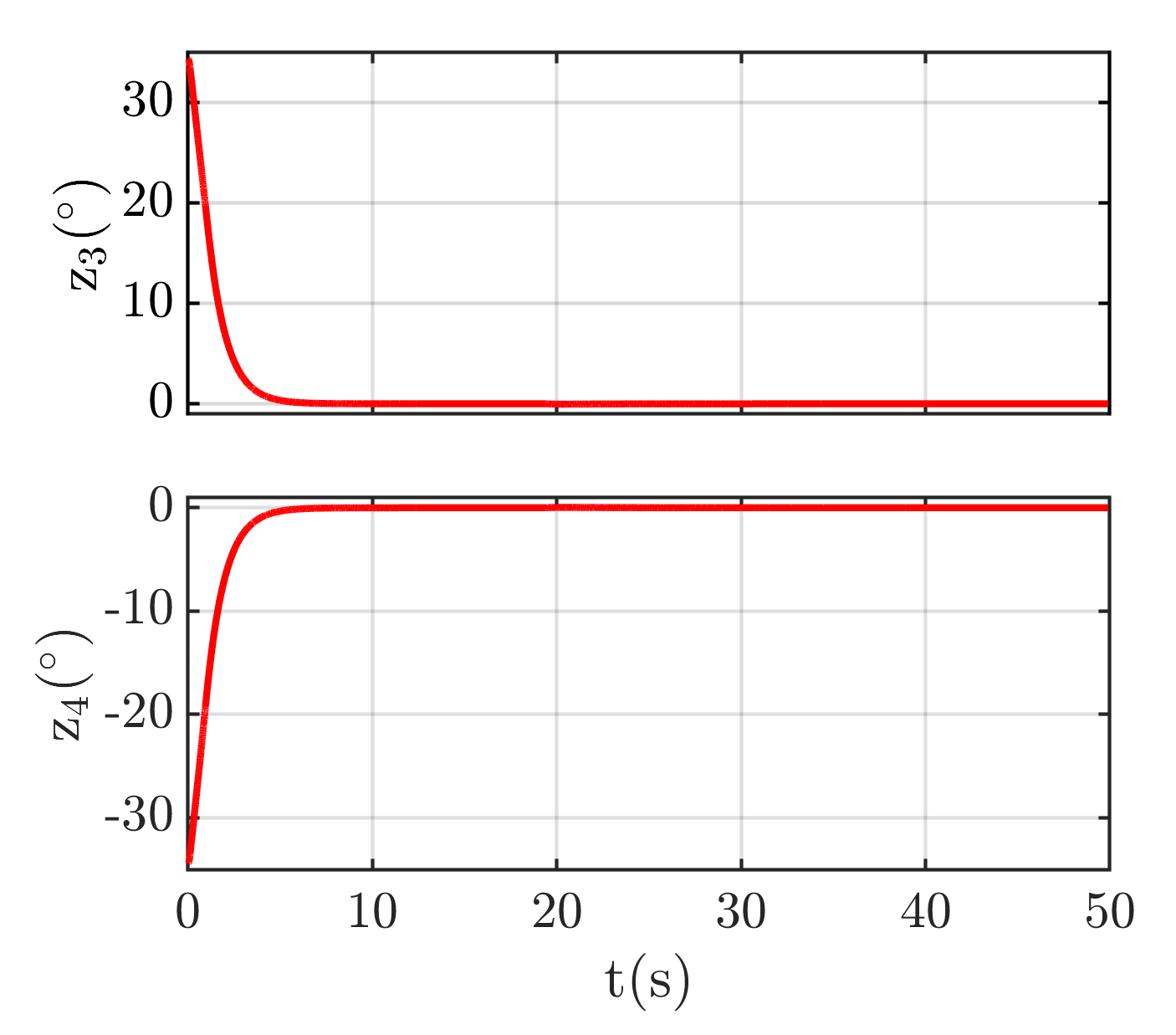}
		\caption{Heading angle errors.}\label{fig:T10_vary_a_max_z3_z4}
	\end{subfigure}
	\begin{subfigure}[b]{0.245\linewidth}
		\includegraphics[width=\linewidth]{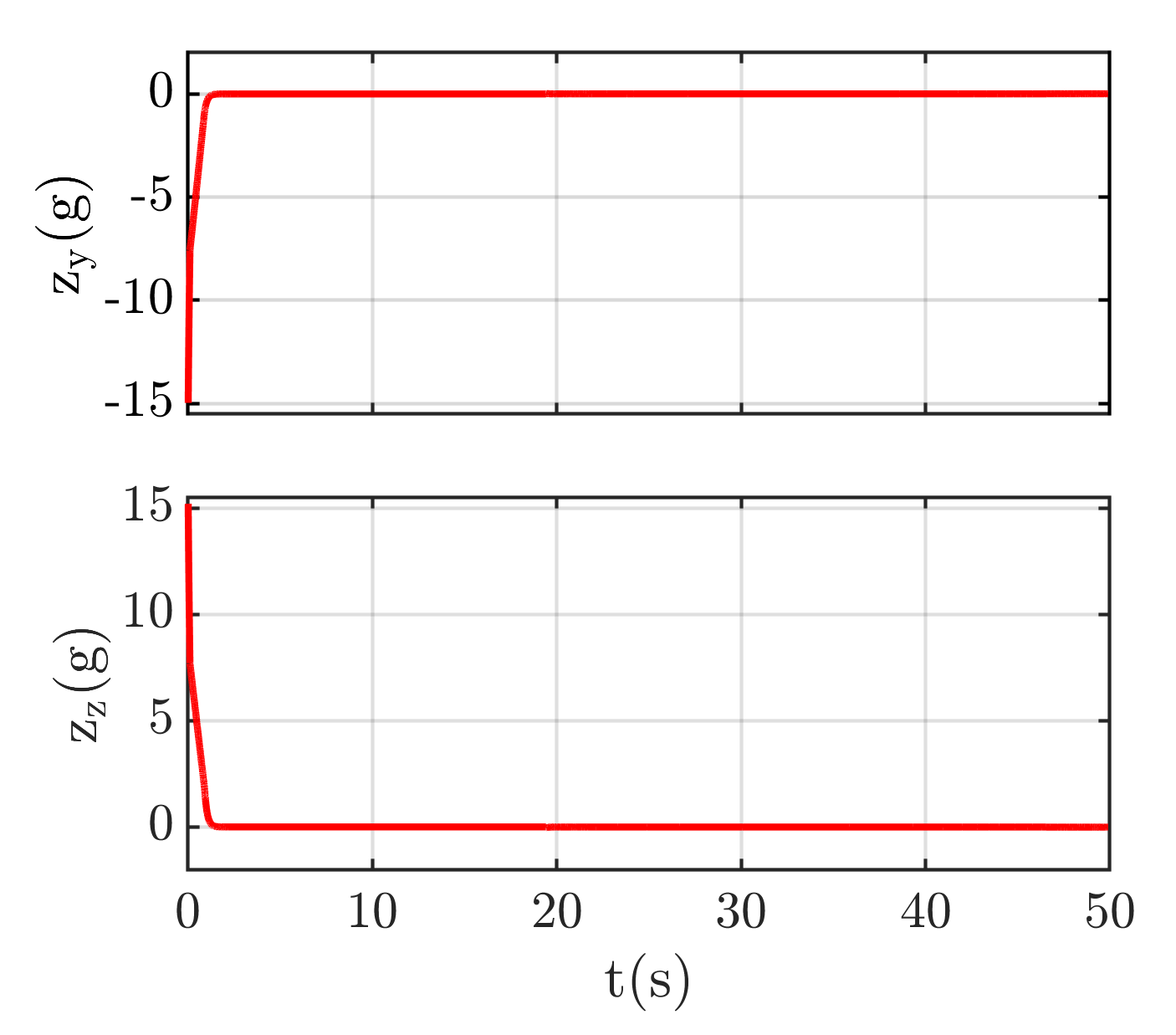}
		\caption{Acceleration errors.}\label{fig:T10_vary_a_max_zy_zz}
	\end{subfigure}
	\caption{Performance study with the time-varying acceleration bounds in \eqref{eq:T10_ayMax_azMax_aMax_1}.}
	\label{fig:T10_vary_a_max}
\end{figure*}
One may observe from \Cref{fig:T10_vary_a_max_ay_az} that the acceleration bound tends to zero when the corresponding acceleration goes to zero. Specifically, when $a_{\rm y}= 0$ and $a_{\rm z}$ is in the vicinity of zero, we have $a_{\rm y,\max}=0$ and $a_{\rm z,\max}=a_{\max}$. This is because of the assumption that the entire lateral acceleration is produced by the wings (or that the interceptor does not have a vertical tail). However, in reality, most interceptors have vertical and horizontal tail fins with a planform area of about $15-25\%$ of the wing area. Therefore, we can assume that some percentage of the feasible acceleration does not depend on the roll of the interceptor. This further leads to the following relations:
\begin{subequations}\label{eq:T10_ayMax_azMax_aMax_2}
	\begin{align}
		a_{\rm y,\max} &= a_{\max,l} + (a_{\max} - a_{\max,l}) \dfrac{a_y}{\sqrt{a^2_{\rm y} + a^2_{\rm z}}},\\
		a_{\rm z,\max} &= a_{\max,l} + (a_{\max} - a_{\max,l}) \dfrac{a_z}{\sqrt{a^2_{\rm y} + a^2_{\rm z}}},
	\end{align}
\end{subequations}
where $a_{\max,l}$ is the fraction of $a_{\max}$ that is independent of the roll position of the interceptor. \Cref{fig:T10_vary_a_max_l} shows the simulation results for the acceleration bounds given in \eqref{eq:T10_ayMax_azMax_aMax_2} with $a_{\max,l}$ and $a_{\max}$ fixed as $1g$ and $5g$ respectively. All other simulation parameters are the same as listed in \Cref{tab:T10_3_Sim_params}.
In \Cref{fig:T10_vary_a_max_l_ay_az}, it may be seen that the acceleration bound no longer becomes zero. The minimum value it can take is $a_{\max,l}=1\,g$. \Cref{fig:T10_vary_a_max_l_ay_az} further indicates that the actual acceleration adheres to its bound even if the commanded acceleration is beyond the feasible range. Once the commanded acceleration enters the feasible range, the actual acceleration converges to the desired acceleration quickly. Decreasing $a_{\max}$ from $10\,$g to $5\,$g introduced a second region of acceleration saturation as inferred by comparing \cref{fig:T10_vary_a_max_ay_az,fig:T10_vary_a_max_l_ay_az}. This explains the peak in the heading angle and acceleration errors as seen in \cref{fig:T10_vary_a_max_z3_z4,fig:T10_vary_a_max_l_zy_zz}. However, the errors later converge once the desired acceleration enters the feasible range. From \Cref{fig:T10_vary_a_max,fig:T10_vary_a_max_l}, one can observe that the impact time constraint is satisfied while adhering to FOV and the acceleration bounds. Moreover, the convergence of the lead angle to zero in \Cref{fig:T10_vary_a_max_r_sigma,fig:T10_vary_a_max_l_r_sigma} implies that the interceptor attains the collision course sufficiently prior to interception.

\subsection{Performance comparison for two-dimensional engagement scenario}
\begin{figure*}[!h]
	\centering
	\begin{subfigure}[b]{0.245\linewidth}
		\includegraphics[width=\linewidth]{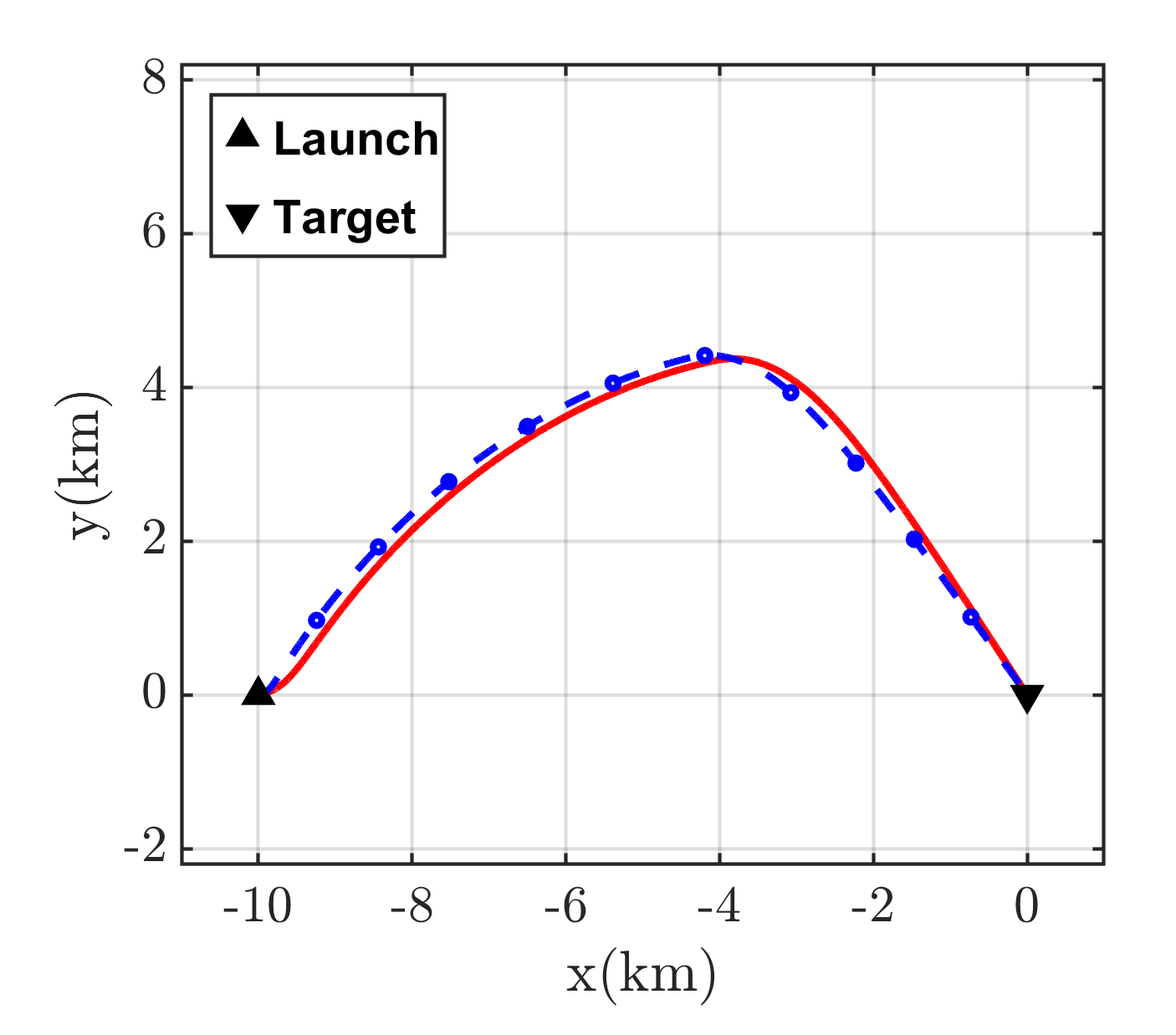}
		\caption{Trajectory}\label{fig:comp_sim_trajectory}
	\end{subfigure}
	\begin{subfigure}[b]{0.245\linewidth}
		\includegraphics[width=\linewidth]{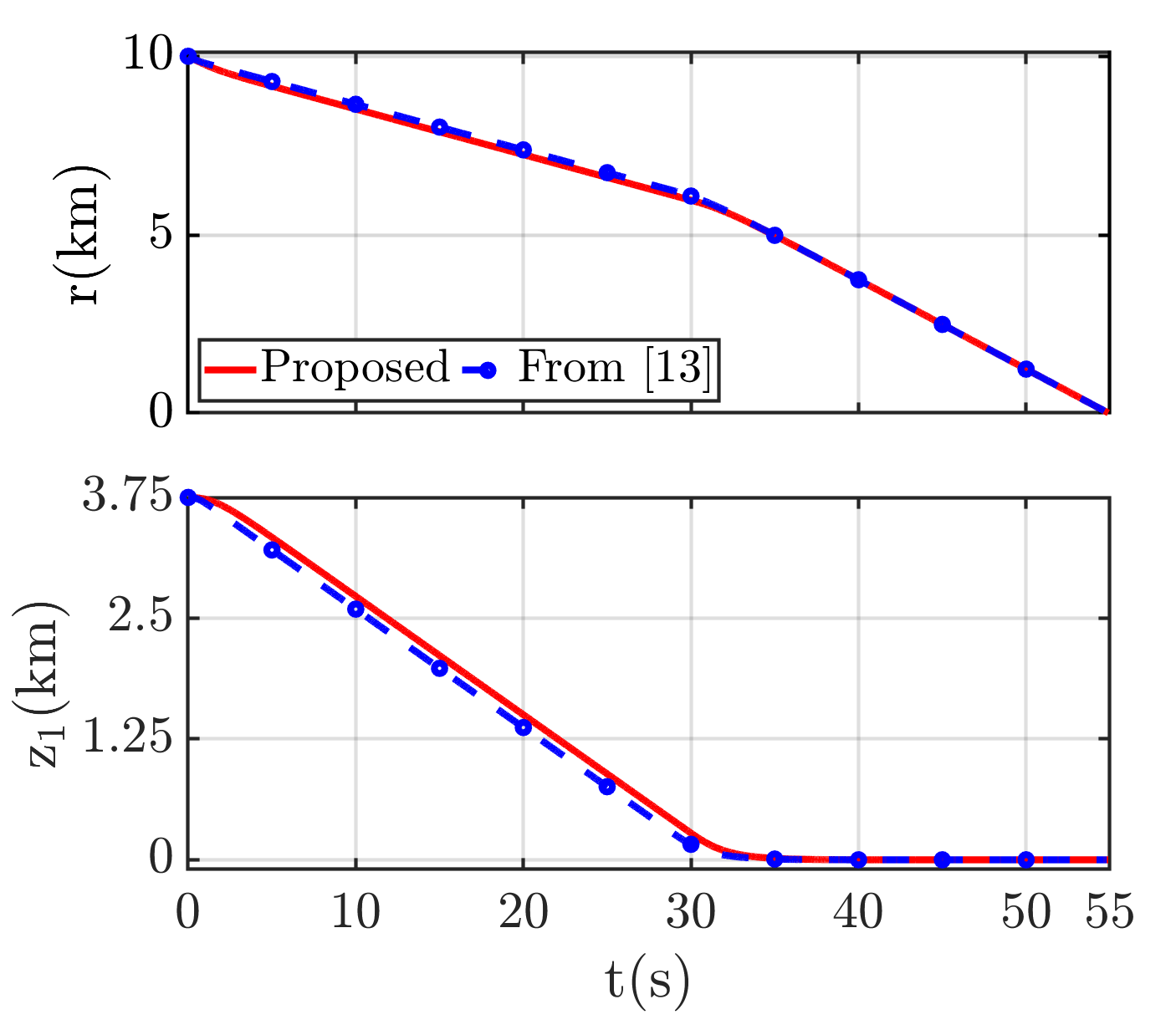}
		\caption{Range and range error}\label{fig:comp_sim_r_z1}
	\end{subfigure}
	\begin{subfigure}[b]{0.245\linewidth}
		\includegraphics[width=\linewidth]{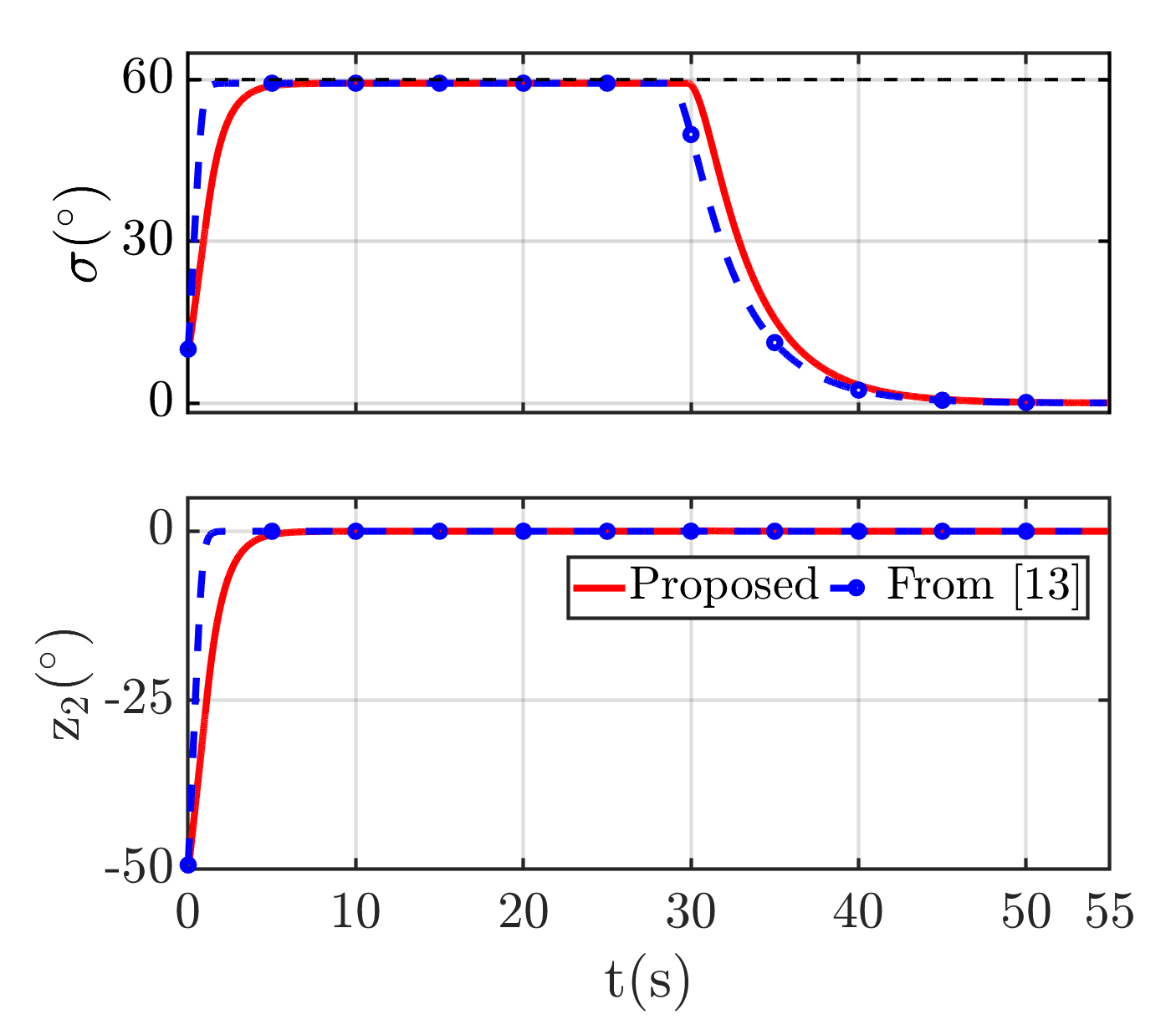}
		\caption{Lead angle and its error}\label{fig:comp_sim_sigma_z2}
	\end{subfigure}
	\begin{subfigure}[b]{0.245\linewidth}
		\includegraphics[width=\linewidth]{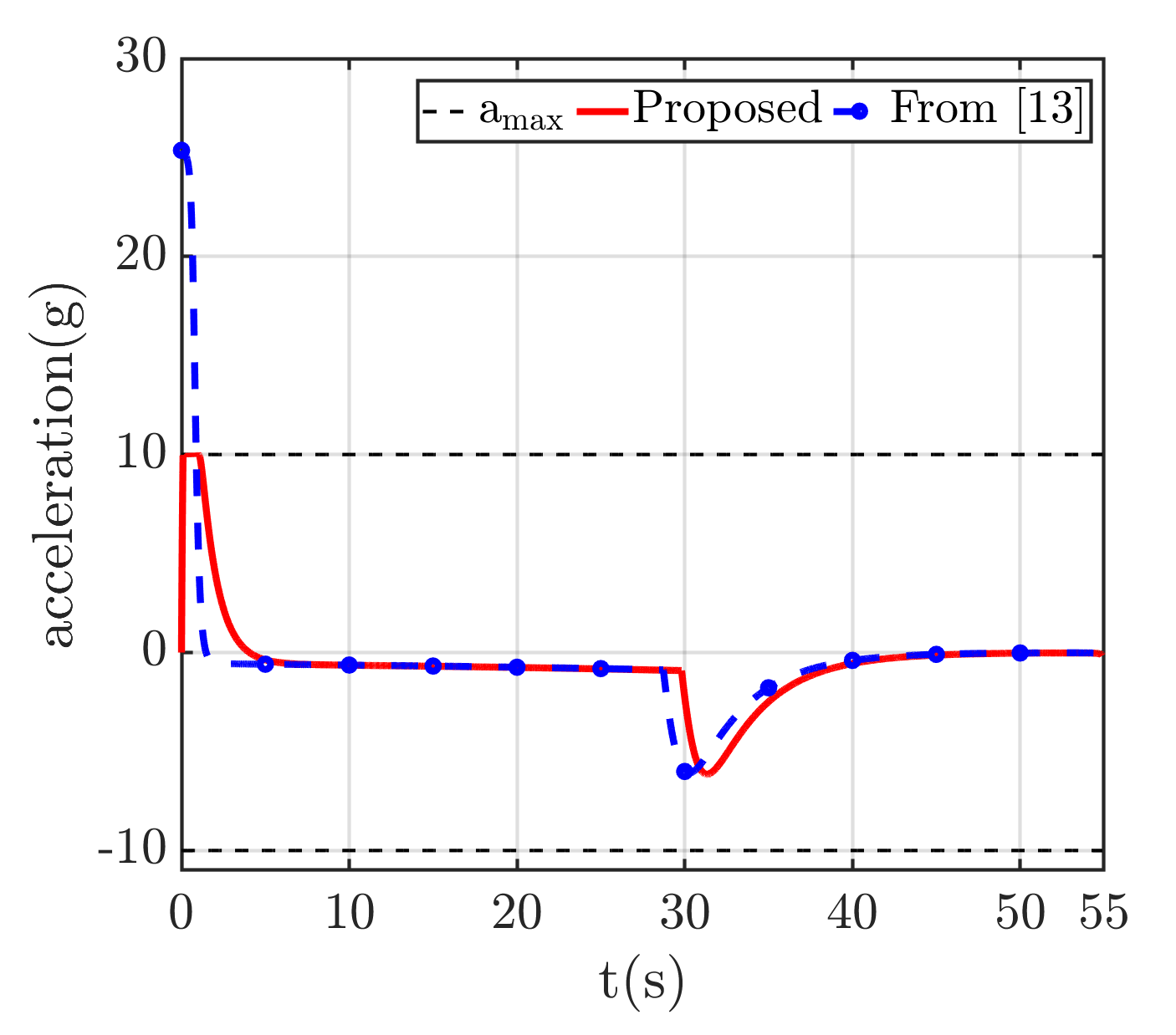}
		\caption{Lateral acceleration}\label{fig:comp_sim_a}
	\end{subfigure}
	\caption{Comparison of with an existing guidance scheme.}
	\label{fig:comp_sim}
\end{figure*}

To further substantiate the efficacy of the proposed strategy, our results are compared with the one in \cite{H_J_Kim}. We use the simulation parameters specified in \Cref{tab:T10_3_Sim_params}. The guidance law design discussed in \cite{H_J_Kim} does not take into account the physical bounds of the actuator. As seen in \Cref{fig:comp_sim}, both the guidance strategies enable the interceptor to successfully intercept the target at the desired impact time.
We observe that the range and lead errors converge faster for the guidance strategy as in \cite{H_J_Kim}, as shown in \Cref{fig:comp_sim_r_z1,fig:comp_sim_sigma_z2}. However, the lateral acceleration demand exceeds its bounds for \cite{H_J_Kim}, whereas it adheres to the constraints for the strategy derived in 
\Cref{thm:ACC_thm1}, as depicted in \Cref{fig:comp_sim_a}. 
Also, the total control effort, defined as $\int a_{\rm M}^2 dt$, is $26321.776\,\rm m^2/s^3$ for the proposed guidance law, whereas for \cite{H_J_Kim}, it is $55281.157\,\rm m^2/s^3$. This further attests to the superiority of the proposed guidance scheme. Additionally, we have also computed the total control effort for different impact times and initial headings. The corresponding values are listed in \Cref{tab:comp_prop_kim}. The proposed guidance law consistently outperforms that in \cite{H_J_Kim}, as indicated by the total control effort values in \Cref{tab:comp_prop_kim}.
\setlength{\tabcolsep}{4pt}
\begin{table}[htpb]
	\centering
	\caption{Comparison of total control effort}
	\begin{tabular}{cccc}
		\hline \hline
		\multirow{2}{*}{Impact time} & \multirow{2}{*}{Initial flight path angle} & \multicolumn{2}{c}{Total control effort} \\
		& & Proposed & From \cite{H_J_Kim}\\
		s & degree & $\rm m^2/s^3$ & $\rm m^2/s^3$\\
		\hline
		$50$ & $10$ & $26453.596$ & $55281.157$ \\
		$50$ & $20$ & $22142.337$ & $44506.869$ \\
		$50$ & $30$ & $17622.982$ & $33871.696$ \\
		$55$ & $10$ & $27745.7$ & $56462.34$ \\
		$60$ & $10$ & $29834.925$ & $58331.555$ \\
		$65$ & $10$ & $33995.707$ & $61802.478$ \\
		\hline\hline
	\end{tabular}
	\label{tab:comp_prop_kim}
\end{table}

\section{Conclusion}
In this paper, we proposed nonlinear guidance strategies for both planar and three-dimensional engagement scenarios to achieve target interception at a desired impact time while satisfying field-of-view (FOV) and input constraints against a stationary target. The proposed guidance schemes remain effective even in scenarios with large initial heading angle errors due to the nonlinear framework used for their derivation. A smooth input saturation model was employed to incorporate lateral acceleration constraints into the guidance design. Furthermore, this saturation model accommodates time-varying acceleration bounds, which are utilized to study the performance of interceptors with different wing-tail lift contributions. Despite the imposed constraints, the interceptor successfully intercepted the target and met the required impact time. Our observations show that, with the proposed guidance scheme, all relevant errors and lateral acceleration converge to zero at the time of interception. Numerical simulations further indicate that the lateral accelerations remain bounded even for very high commanded input values. Finally, we compared the performance of the proposed guidance strategy with an existing method, and the proposed strategy demonstrated superior performance.

\ifCLASSOPTIONcaptionsoff
\newpage
\fi

\bibliographystyle{IEEEtaes}
\bibliography{RefACC}
\end{document}